\documentclass{article}

\usepackage{PRIMEarxiv}

\usepackage{lineno,hyperref}
\usepackage{graphicx}
\usepackage{amsmath}
\usepackage{amsfonts}
\usepackage{amssymb}
\usepackage{bm}
\usepackage{subcaption}
\usepackage{multirow}
\usepackage{algorithm}
\PassOptionsToPackage{noend}{algpseudocode}
\usepackage{algpseudocode}

\usepackage{geometry}
\errorcontextlines\maxdimen

\usepackage[utf8]{inputenc} 
\usepackage[T1]{fontenc}    
\usepackage{hyperref}       
\usepackage{url}            
\usepackage{booktabs}       
\usepackage{nicefrac}       
\usepackage{microtype}      
\usepackage{fancyhdr}       

\title{Model-agnostic stochastic predictive control using limited and output only measurements}

\author{Tapas Tripura\\
  Department of Applied Mechanics\\
  Indian Institute of Technology Delhi\\
  \texttt{tapas.t@am.iitd.ac.in} \\
  \And
      Souvik Chakraborty \\
  Department of Applied Mechanics\\
  Yardi School of Artificial Intelligence (ScAI)\\
  Indian Institute of Technology Delhi\\
  \texttt{souvik@am.iitd.ac.in} \\
}

\begin{document}
\maketitle

\begin{abstract}
We propose a model-agnostic stochastic predictive control (MASMPC) algorithm for dynamical systems. The proposed approach first discovers \textit{interpretable} governing differential equations from data using a novel algorithm and blends it  with a model predictive control algorithm. One salient feature of the proposed approach resides in the fact that it requires no input measurement (external excitation); the unknown excitation is instead treated as white noise, and a stochastic differential equation corresponding to the underlying system is identified. With the novel stochastic differential equation discovery framework, the proposed approach is able to generalize; this eliminates the repeated retraining phase -- a major bottleneck with other machine learning-based model agnostic control algorithms. 
Overall, the proposed MASMPC  (a) is robust against measurement noise, (b) works with sparse measurements, (c) can tackle set-point changes, (d) works with multiple control variables, and (e) can incorporate dead time. We have obtained state-of-the-art results on several benchmark examples. Finally, we use the proposed approach for vibration mitigation of a 76-storey building under seismic loading.
\end{abstract}


\section{Introduction}
Control is an inextricable part of the modern automation process and industrial revolution. The goal of dynamical system control is to devise a strategy for driving an uncontrollable system to the desired state while maintaining control stability. There exist different forms of control, such as passive control, semi-active control, and active control. Passive control mechanisms have low maintenance requirements and are advantageous over active ones since they do not require external energy to operate \cite{dinh2015passive}. However, due to limitations in the control's adaptability, they do not function with the same efficiency for other types of dynamic loading. The semi-active and active control systems, on the other hand, allow effective control with little or no human oversight \cite{du2007adaptive,xu2003semi}. As one of the most successful and popular algorithms in active and semi-active control, model predictive control (MPC) algorithms have achieved a great impact on applications in all areas of control systems \cite{zhou1998essentials,allgower1999nonlinear,mayne2014model}. 

MPC \cite{mayne2014model} is one of the most powerful and advanced model-based forms of optimal control \cite{zhou1998essentials}. Its merits and popularity stem from its flexibility in cost function construction, the ability to incorporate constraints, the power to handle multiple operating constraints, and extensions to nonlinear systems \cite{allgower1999nonlinear}. However, to obtain the requisite corrective behavior, a controller designed using the MPC requires knowledge about the governing physics of the underlying process. However, in practice, we often do not have access to the governing physics of the system. Under such circumstances, the MPC algorithms often rely on empirical models for distilling the physics from data \cite{ozkan2001nonlinear,shehu2016applications}. Unfortunately, empirical models are often not accurate and only yield accurate results in the vicinity of the data from which it is derived. To circumvent this issue, research often relies on advanced data-driven techniques such as eigensystem realization algorithm (ERA) \cite{juang1985eigensystem}, subspace identification methods \cite{hale2002subspace}, singular value decomposition (SVD) \cite{flores2005latent}, recursive canonical correlation analysis (RCCA) \cite{bhowmik2020real,panda2021first}, Kalman filter models \cite{lee1994extended}, autoregressive models \cite{billings2013nonlinear}, neural networks \cite{wang2015combined,zhang2016learning} etc. We note that despite the success of such data-driven techniques, these models generally do not generalize to unseen environments. 

Based on the discussion above, it is evident that the major bottleneck in MPC stems from the unavailability of the governing physics of the system. Physics-informed deep learning (PIDL) \cite{goswami2020transfer,raissi2019physics,guo2020stochastic,beltran2022physics}, operator learning \cite{lu2021learning,garg2022assessment,tripura2022wavelet}, and other machine learning algorithms \cite{sudret2015sparse} have recently emerged as viable alternatives; however, both PIDL and operator learning algorithms are not suitable for online learning. The sparse identification of nonlinear dynamics (SINDy) algorithm proposed in \cite{brunton2016discovering,kaiser2018sparse} is a possible alternative as it learns governing physics from data. A major issue, however, resides in the fact that it requires the external input force information to identify the governing physics correctly. In an online scenario, often large data sets and high training time is not a luxury; in addition, external disturbances are not always possible to measure. When both the data and the input disturbance information are restricted, these challenges limit the online implementation of existing data-driven model learning algorithms for real-time control of high-dimensional nonlinear dynamical systems. In summary, we need a model that is accurate, efficient, and can function with only the output measurements.

To address the challenges discussed above, we herein propose a model-agnostic stochastic model predictive control (MASMPC) algorithm by combing the stochastic differential equations (SDEs), the sparse Bayesian regression, and the stochastic model predictive control (SMPC). The proposed algorithm firstly learns the governing physics entirely from the systems output response without requiring the systems input information and, secondly, uses the discovered physics in the purview of SMPC to control the underlying system. We borrowed the names (i) model-agnostic since the actual information regarding the governing model is initially not available to the proposed approach, instead discovered from the data, and (ii) stochastic because the input information is often not measurable and thus in the proposed problem statement the discovery is targeted with no information on input.

\begin{figure}[t]
    \begin{center}
    \includegraphics[width=\textwidth]{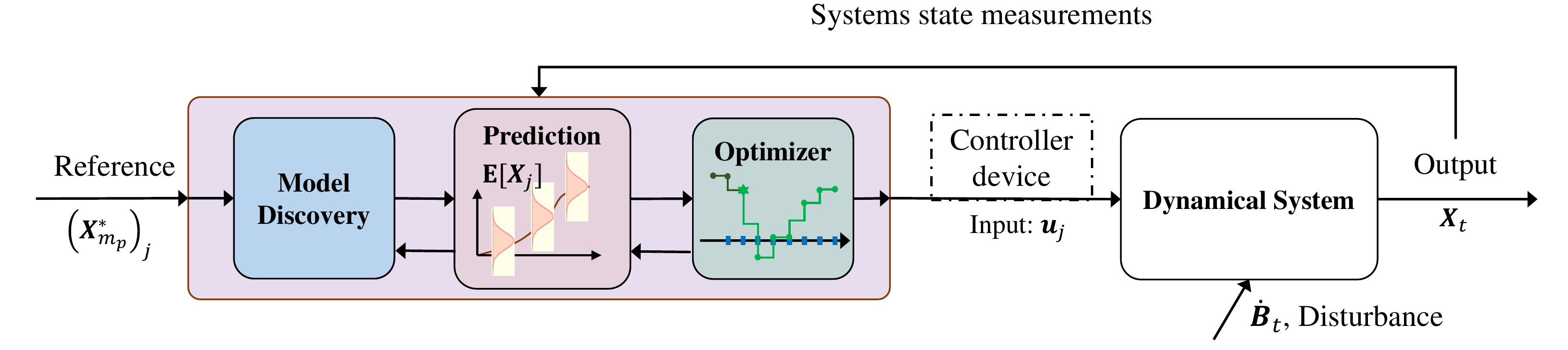}
    \caption{Schematic of the proposed nonlinear model agnostic stochastic model predictive control}
    \label{fig_framework}
    \end{center}
\end{figure}

The proposed MASMPC algorithm is capable of taking into the effect of actuation and able to identify interpretable controlled physics while avoiding overfitting in low-data environments without the use of external input disturbance information. In the restricted knowledge, the external input disturbance is handled as a stochastic process, and the problem is formulated as a stochastic model discovery problem with SMPC \cite{dai2015cooperative,ning2021online,chen2021stochastic}. In cases of non-measurable inputs, the use of SDEs for creating the influence of stochastic input on system dynamics is very common \cite{kloeden1992higher,oksendal2013stochastic}. Thus in the proposed MASMPC, it is assumed that the underlying models are recoverable through stochastic differential equations (SDEs). The Kramers-Moyal expansion is used to express the SDEs in terms of the systems output state measurements \cite{risken1996fokker}. This is followed by sparse Bayesian regression to discover the interpretable governing physics of the underlying process \cite{nayek2021spike}. Because the knowledge about external disturbance is unavailable, the control must be calculated while accounting for the uncertainty and recursive feasibility. This necessitates the use of SMPC \cite{hokayem2012stochastic,dai2015cooperative,di2013stochastic}. We note that SMPC results in an optimal control that is less conservative than the traditional MPC algorithms. The overarching framework of the proposed data-driven nonlinear model predictive control is shown in Fig. \ref{fig_framework}.
The salient features of the proposed MASMPC can be encapsulated in the following points:
\begin{itemize}
    \item \textbf{Robustness against measurement noise}: The proposed framework adopts a Bayesian approach and hence, is less prone to overfitting. This results in a \textit{robust} control algorithm that can handle noisy measurement in a robust manner. This is illustrated later in Section \ref{sec:result}.
    \item \textbf{Generalization}: 
    Unlike other model agnostic control algorithms, the proposed approach does not rely on black-box approaches; instead, it first learns the governing SDE that best explains the data. Overall, such a  setup results in a model that is interpretable and generalizes to a new environment. This is particularly important in the context of model-agnostic control algorithms, as the repeated retraining of the learned model can be avoided. We have illustrated in Section \ref{sec:result} that the proposed algorithm trained with only $1$s of data yields satisfactory results perpetually. 
    \item \textbf{Time varying set-points, multiple control variables, and dead-time}: The proposed MASMPC algorithm is mathematically well equipped to handle the variation in set-points and provides a readymade platform to control systems with more than one control variable. Additionally, the proposed approach can also incorporate dead time because of the transmission of control action and corresponding reaction.
    These are also illustrated in Section \ref{sec:result}.
\end{itemize}

The remainder of this paper is organized as follows: in section \ref{sec:pf}, the proposed framework for simultaneous interpretable model learning and stochastic model predictive control is illustrated. In section \ref{sec:result}, the main results are demonstrated. In section \ref{sec:conclusion}, the salient features of the proposed framework are reviewed, and then the paper is concluded.

\section{Proposed framework: Stochastic model predictive control with learning}\label{sec:pf}
The proposed framework integrates the concepts of physics discovery of dynamical systems with the model predictive control in the absence of the input measurements; in order to achieve faster model learning and exact control of nonlinear dynamical systems while maintaining minimum dependency on data. Before proceeding to the construction of the framework, let us define the complete probability space by $\left( \Omega, \mathcal{F}, \mathcal{P} \right)$, where $\mathcal{F}$ is the $\sigma$-algebra constructed from the subsets from $\Omega$ and $\mathcal{P}: \mathcal{F} \mapsto [0, 1]$ is a probability measure on the measurable space $\left( \Omega, \mathcal{F} \right)$. Let $E$ be an arbitrary non empty space, then for each $t \in T$, where $T$ is another arbitrary set, the random variable $X$ defines a mapping  $X: \omega \mapsto E$ such that there exists an inverse map ${X^{-1}}: \{\mathcal{E}\in E \} \mapsto \{\omega \in \Omega\} $. The collection of random variables $\left\{X_t : t \in T \right\}$ on $\left(\Omega, \mathcal{F} \right)$ constructs a stochastic process. To formulate the framework of the proposed MASMPC approach, let us consider the following dynamical model,
\begin{equation}\label{odeg}
    \dot{\bm{X}}_t = F(\bm{X}_t,\bm{u}_t,t); \quad \bm{X}(t=t_0)=\bm{X}_0,
\end{equation}
where ${\bm{X}}_t \in {\mathbb{R}^m}$ denotes the ${{\mathcal{F}_t}}$-measurable state vector, $\bm{u}_t \in \mathbb{R}^{q}$ is the control vector, and $F : \mathbb{R}^m \times \mathbb{R}^q \mapsto \mathbb{R}^m$ represents nonlinear but smooth and well-behaved function defining the dynamics of the underlying process. The above model is valid when the complete knowledge about the inputs and outputs of the system is known. However, in most cases obtaining the input measurements are intractable. In such cases, the input forces in the model can be modeled as stochastic forces, this transfers the deterministic model discovery to a stochastic domain. The dynamics in Eq. \eqref{odeg} are thus represented in terms of the stochastic differential equations (SDEs). Details on SDE are provided in the next section.

\subsection{Sparse learning of Stochastic differential equations}
SDE provides a splendid approach for describing the dynamics of non-linear stochastic dynamical systems subjected to random forces \cite{kloeden1992higher}. An SDE takes into account the deterministic dynamics through its drift component, and the random forces are presented through the additive stochastic diffusion components \cite{kloeden1992higher,tripura2020ito}. If $\{{\mathcal{F}_t},0 \le t \le T)\}$ is the natural filtration constructed from sub $\sigma$-algebras of $\mathcal{F}$, then under the complete probability space $\left( {\Omega ,\mathcal{F}, \mathcal{P}} \right)$ an $m$-dimensional $n$-factor SDE is defined as \cite{tripura2020ito,tripura2021change}, 
\begin{equation}\label{sdeg}
    \begin{aligned}
        &d{{\bm{X}}_t} = {\bm f}\left( {{{\bm {X}}_t}, {\bm{u}_t},t} \right)dt + {\bm g}\left( {{{\bm {X}}_t},{\bm{u}_t},t} \right)d{\bm B}\left( t \right); {\bm X}(t=t_0)={\bm X}_0; t \in [0,T],
    \end{aligned}
\end{equation}
where ${\bm{f}}\left( {{{\bm{X}}_t},{\bm{u}_t},t} \right):{\mathbb{R}^m} \mapsto {\mathbb{R}^m}$ is the drift vector, ${\bm{g}}\left( {{{\bm{X}}_t},{\bm{u}_t},t} \right):{\mathbb{R}^m} \mapsto {{\mathbb{R}}^{m \times n}}$ is the diffusion matrix, and \{${\bm{B}}_j(t),j = 1, \ldots n$\} is the ${\mathbb{R}^n}$ Brownian motion vector. The solution to Eq. \eqref{sdeg} can be obtained using various stochastic integration schemes \cite{kloeden1992higher,tripura2020ito,tripura2021change}. In general the stochastic Brownian motion $B(t)$ is not a well defined mathematical function; however, under the Lipschitz continuity the SDE components ${\bm f}\left( {{{\bm {X}}_t},{\bm{u}_t},t} \right)$ and ${\bm g}\left( {{{\bm {X}}_t},{\bm{u}_t},t} \right)$ constructs a unique solution to the SDE in Eq. \eqref{sdeg} \cite{tripura2020ito}. The drift ${\bm f}\left( {{{\bm {X}}_t},{\bm{u}_t},t} \right)$ is generally a well-behaved deterministic function with a finite variation. On the contrary, the Brownian motions are non-differentiable functions that have zero finite variations but non-vanishing quadratic variations. In such cases, the drift and diffusion components can be expressed in terms of the linear and quadratic variations of the measurement time history data by using the Kramers-Moyal expansion as (the detailed derivation of the expansion is presented in \ref{appenA}):
\begin{equation}\label{kramers}
    \begin{aligned}
    {f_i}\left( {{{\bm {X}}_t},{\bm{u}_t},t} \right) &= {\left. {\mathop {\lim }\limits_{\Delta t \to 0} \frac{1}{{\Delta t}}E\left[ {{X_i}(t + \Delta t) - {\xi_i}} \right]} \right|_{{x_k}(t) = {\xi _k}}} \forall \;k = 1, \ldots N\\
    {\Gamma_{ij}}\left( {{{\bm {X}}_t},{\bm{u}_t},t} \right) &= \frac{1}{2}\mathop {\lim }\limits_{\Delta t \to 0} \frac{1}{{\Delta t}}E\Bigl[ \left| {{X_i}(t + \Delta t) - {\xi_i}} \right| {\left. { \left| {{X_j}(t + \Delta t) - {\xi_j}} \right| \Bigr]} \right|_{{x_k}(t) = {\xi_k}}}\forall \;k = 1, \ldots N,
    \end{aligned}
\end{equation}
where ${f_i}\left( {{{\bm {X}}_t},{\bm{u}_t},t} \right)$ is the $i^{th}$ drift component and ${{\bf{\Gamma}}_{ij}}$ is the $(ij)^{th}$ component of the diffusion covaraince matrix ${\bf{\Gamma}} \in {\mathbb{R}^{m \times m}} := g({{\bm{X}}_t},{\bm{u}_t},t)g{({{\bm{X}}_t},{\bm{u}_t},t)^T}$. The discovery of the governing SDE requires expressing the drift and diffusion components in a combination of the candidate basis functions. The identification of the candidate function that best represents the data can be performed through sparse regression. The idea here is to recover a model of the governing process having only a few key functions, thereby rendering the model to be interpretable and accurate enough to perform future predictions with minimum possible error. Towards this, a regression problem for learning the drift and diffusion terms can be formulated as,
\begin{subequations}\label{regress}
    \begin{align}
        {{\bm{Y}}_i} &= {\bf{L}}{{\bm{\theta }}_i^{f}
        } + {{\bm{\varepsilon }}_i} \label{regres_drift} \\
        {{\bm{Y}}_{ij}} &= {\bf{L}}{{\bm{\theta }}_{ij}^{g}} + {{\bm{\eta }}_i}. \label{regres_diff}
    \end{align}
\end{subequations}
Eq. \eqref{regres_drift} corresponds to the regression problem for the drift component, where the target vector ${\bm{Y}}_i$ is constructed from the linear variations of the measurement path of the $i^{th}$ degree-of-freedom (DOF) as, ${{\bm{Y}}_i} = {\left[ {\left( {{X_{i_1}} - {\xi _{i_1}}} \right), \ldots, \left( {{X_{i_N}} - {\xi _{i_N}}} \right)} \right]^T}$. Similarly, Eq. \eqref{regres_diff} defines the regression problem for the diffusion component, where the target vector is set as, ${{\bm{Y}}_{ij}} = {\left[ {\left( {{X_{i_1}} - {\xi _{i_1}}} \right)\left( {{X_{j_1}} - {\xi _{j_1}}} \right), \ldots, \left( {{X_{i_N}} - {\xi _{i_N}}} \right)\left( {{X_{j_N}} - {\xi _{j_N}}} \right)} \right]^T}$. This corresponds to the quadratic covariation of the measurement paths of the $i^{th}$ and $j^{th}$ DOF. In both cases, ${X_{i_j}}$ denotes one step ahead point of ${\xi _{i_j}}$. The drift and diffusion weight vectors are defined as, ${{\bm{\theta }}_i^{f}} = {\left[ {{\theta _{i_1}}},{{\theta _{i_2}}}, \ldots ,{{\theta _{i_K}}} \right]^T}$ and ${\bm{\theta }}_{ij}^g = {\left[ {\theta _{{ij}_1}},{\theta _{{ij}_2}}, \ldots ,{\theta _{{ij}_K}} \right]^T}$. In an $m$-dimensional diffusion process, each of the drift terms can be expressed in terms of the candidate basis functions as ${f_i}({{\bm{X}}_t},t) = \sum\nolimits_{k = 1}^K {{\upsilon _k}({{\bm{X}}_t}){\theta _{i_k}^{f}}}$; $i=1, \ldots, m$ where, $\{ {\upsilon _k}({{\bm{X}}_t});\;k = 1,2, \ldots, K \}$ denotes the candidate basis functions and $\{ {\theta _{i_k}}; k=1, \ldots, K \}$ denotes the weight vector corresponding to the $i^{th}$ drift component. Similarly, the $(ij)^{th}$ component of the diffusion covariance matrix ${\bf{\Gamma}}$ can be expressed in terms of the candidate basis functions as ${{\bf{\Gamma }}_{ij}} = \sum\nolimits_{k = 1}^K {{\upsilon _k}({{\bf{X}}_t}){{\bf{\theta }}_{{ij}_k}^{g}}}$; $i,j = 1,2, \ldots, m$. Therefore the library of candidate functions in Eq. \eqref{regress} can be constructed from the candidate basis functions as, 
\begin{equation}
	{\bf{L}}({\bm{X}, \bm{u}}) = \left[ {\begin{array}{*{20}{c}}
			{\bf{1}} &{{\upsilon _k}({\bm{X}})} &{{\upsilon _k}({\bm{u})}} &{{\upsilon _k}({{\bm{X}} \otimes {\bm{X}}})}&{{\upsilon _k}({{\bm{X}} \otimes \bm{u}})}& \ldots \end{array}} \right],
\end{equation}
where ${{\bm{X}}_t} \otimes {\bm{u}}$ represents the combinations of the state and control force vector. Once the structure for the linear regression is formulated, a straightforward application of the sparse Bayesian regression, discussed in the next section, would provide the governing model of the underlying process.
\begin{figure}[t]
    \begin{center}
    \includegraphics[width=\textwidth]{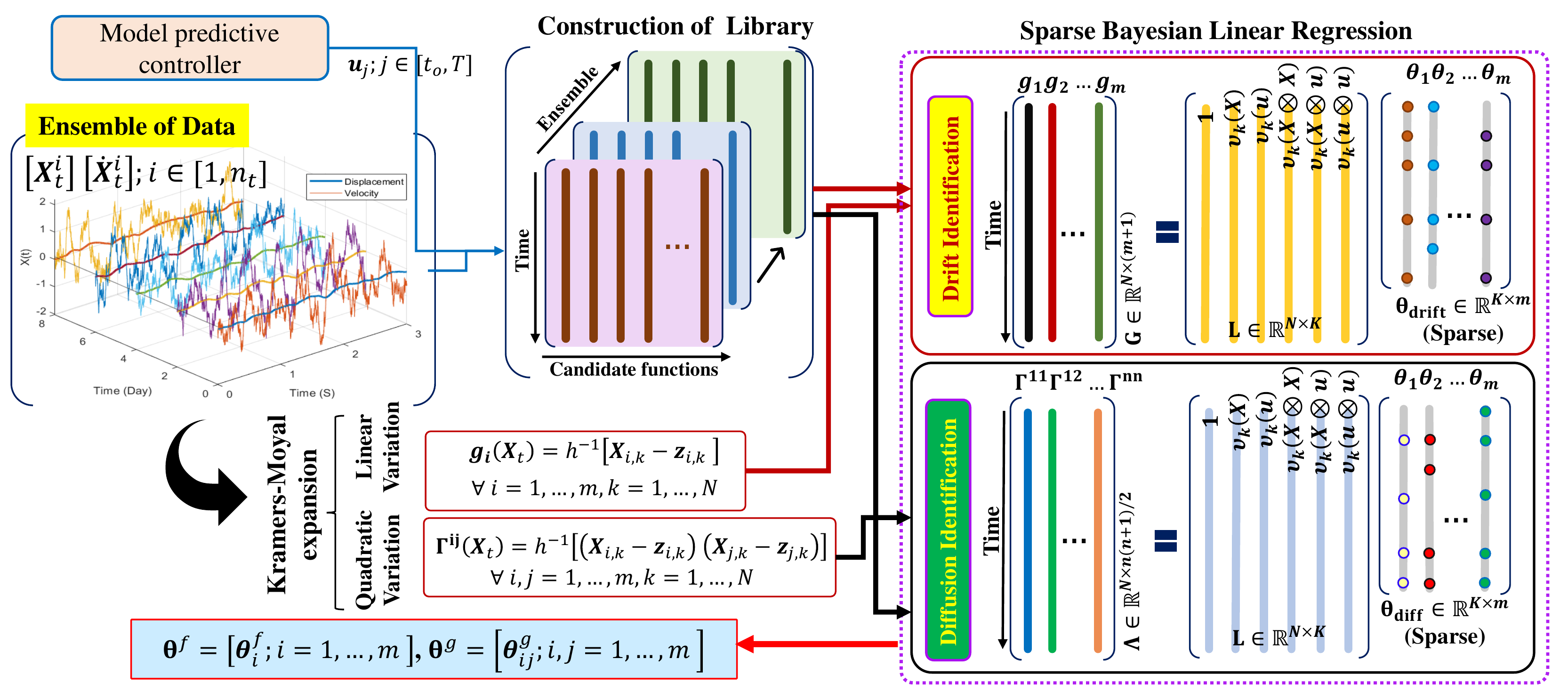}     
    \caption{Schematic of the proposed Bayesian-SDE framework for model discovery without input measurement. The first step in the framework is the construction of the target vector using the Kramers-Moyal expansion. In the second step, given an ensemble of the time history of data and the control force, the framework obtains the expectation of the sequence of libraries. The candidate function in the library is parameterized by a weight vector. The drift and diffusion terms are then discovered using sparse Bayesian linear regression.}
    \label{fig_SINDy}           
    \end{center}     
\end{figure}
Noting that an $m$-dimensional SDE has $m$-dimensional drift vector, it requires $m$ regression defined in Eq. \eqref{regres_drift} to be performed for identifying all the drift components. The covariance matrix of the $m \times n$-diffusion matrix has a dimension $m \times m$. But the covariance matrix is symmetric in nature; therefore, one needs to perform ${{m(m + 1)}}/{2}$ regression in Eq. \eqref{regres_diff} to completely characterize the diffusion space. In Fig. \ref{fig_SINDy}, the approach of learning SDEs without the measurements of input disturbance is demonstrated.

\subsection{Discovery of SDE by sparse Bayesian regression}\label{section21}
For performing the regressions defined in Eq. \eqref{regress} using Bayesian inference, let us assume that ${\bf{L}} \in \mathbb{R}^{N \times K}$ is the library of candidate functions, composed from $K$ independent basis functions. The term $N$ here denotes the length of the sample paths. The main idea of the sparse Bayesian linear regression is to identify $k$ basis functions out of the total $K$ candidate library functions such that, as a linear combination, these $k$ bases represent the sample path accurately. The coefficients of the candidate function in the linear combination are denoted by the weight vector ${\bm{\theta}}$. In general, $k \ll K$, thereby resulting sparse solution. The condition $k \ll K$ is maintained by assigning sparsity promoting prior on the weight vector ${\bm{\theta}}$. Among the alternative candidates, the spike and slab (SS) prior have shown high shrinkage properties due to its sharp spike at zero and a diffused density spanned over a large range of possible parameter values \cite{george1997approaches,o2009review}. There exist different versions of spike and slab distribution; however, we model the spike at zero as Dirac-delta function and the tail distribution as independent Student's-t distribution; this is referred to as DSS-prior though out the text. The Dirac-delta helps in concentrating most of the probability mass at zero, while the diffused tail distributes only a small amount of probability mass over a large range of possible values. Thus, the DSS-prior allows only a few samples with a very high probability to escape shrinkage. 

\begin{figure}[t]
	\centering
	\includegraphics[width=0.3\textwidth]{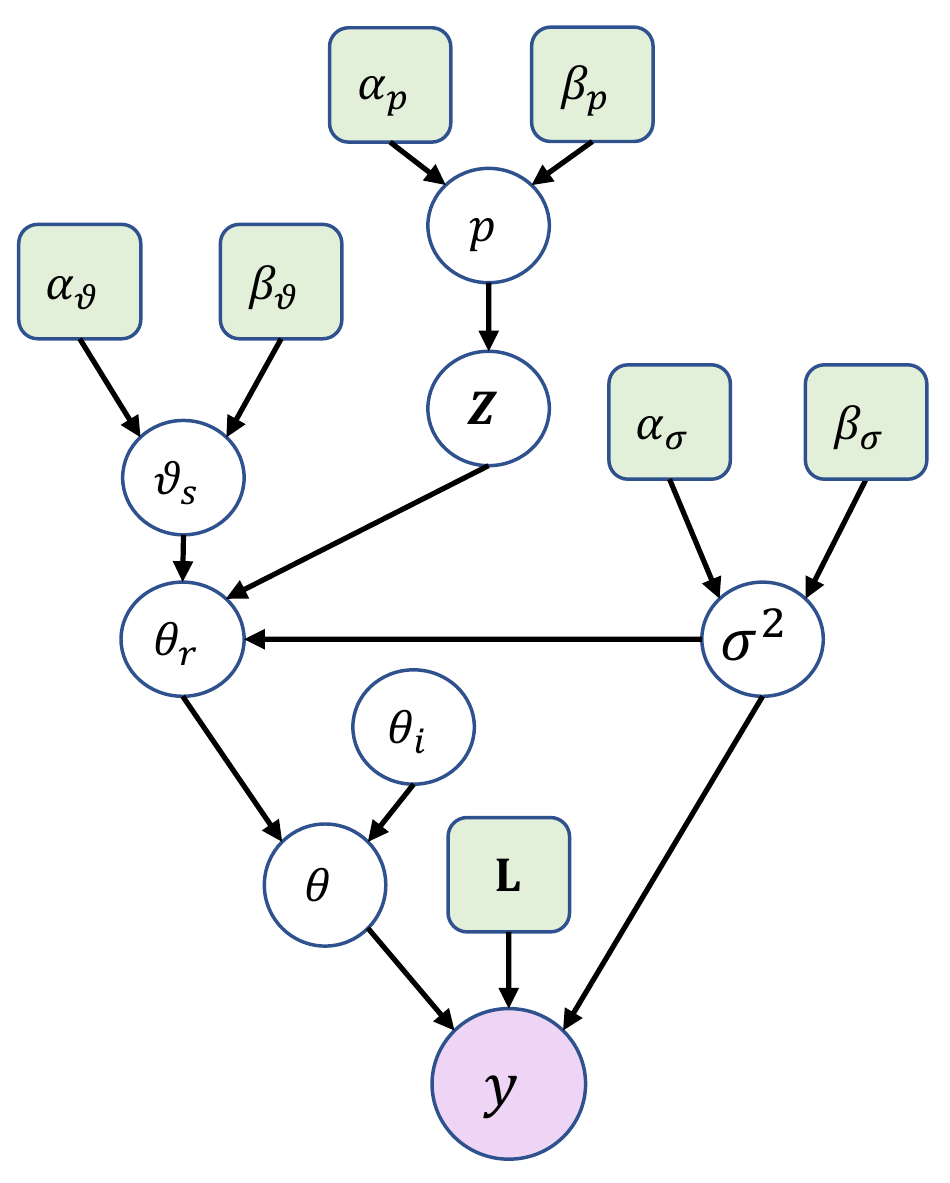}
	\caption{Directed Acyclic Graph of the discontinuous spike and slab model for sparse Bayesian linear regression. The variables in the green square boxes indicate the deterministic parameters, and the variables in the white circles represent random variables. The term ${\bf{L}}$ represents the library of candidate functions, and the parameters ${\alpha _\vartheta }$, ${\beta _\vartheta }$, ${\alpha _p }$, ${\beta _p }$, ${\alpha _\sigma }$, and ${\beta _\sigma }$ indicates the hyperparameters of the priors of the hierarchical DSS model. Here $p$, $\vartheta_s$ and $\sigma^2$ scalar-valued, and, the the variables ${\bm{Z}}$ and ${\bm{\theta}}$ are vector-valued variables.}
	\label{fig_graph}
\end{figure}

It is therefore, easy to understand that the weights associated with the functions which do not participate in the representation of the data belong to spike, and those which have contributions to the presentation of data belong to slab. Classification of the weights into spike and slab components of the DSS-prior is done by introducing a latent indicator variable ${\bm{Z}}=\left[ Z_1, \ldots, Z_K \right]$ for each of the component $\theta_k; k = 1, \ldots, K$ in the weight vector ${\bm{\theta}}$. The variable $Z_k$ takes a value of 1 if the weight $i$ corresponds to the slab component; otherwise, it takes a value of 0 when the weight belongs to spike component. Let us represent the Eq. \eqref{regress} by following one-dimensional regression equation,
\begin{equation}\label{regression}
    {\bm{Y}} = {\bf{L}}{\bm{\theta}} + \epsilon,
\end{equation}
where ${\bm{Y}} \in \mathbb{R}^N$ denotes the $N$-dimensional target vector, ${\bf{L}}$ denotes the library of candidate functions, ${\bm{\theta}}$ is the weight vector, and $\epsilon \in \mathbb{R}^N$ is the residual error vector representing the model mismatch error. The model error $\epsilon$ is modeled as i.i.d Gaussian random variable with zero mean and variance $\sigma^2$. The application of the Bayes formula for estimating the weights yields,
\begin{equation}
    P\left( {\bm{\theta} |{\bm{Y}}} \right) = \frac{{P\left( {{\bm{Y}}|{\bm{\theta}} } \right)}{{P\left( {\bm{\theta}} \right)}}}{{P\left( {\bm{Y}} \right)}},
\end{equation}
where ${P\left( {\bm{\theta}} \right)}$ is the prior distribution on the weight vector and ${P\left( {{\bm{Y}}|{\bm{\theta}} } \right)}$ is the likelihood function. The likelihood function is given as,
\begin{equation}
    {\bm{Y}}|{\bm{\theta }},{\sigma ^2} \sim \mathcal{N}\left( {{\bf{L}}{\bm{\theta}} ,{\sigma ^2}{{\bf{I}}_{N \times N}}} \right),
\end{equation}
where ${\bf{I}}_{N \times N}$ denotes the ${N \times N}$ identity matrix. For the present problem, the prior ${P\left( {\bm{\theta}} \right)}$ is chosen as the spike and slab distribution. It is to be noted that in the estimate of the posterior distribution $P\left( {\bm{\theta} |{\bm{Y}}} \right)$, only the components of the weight vector that belongs to the slab contribute. Thus, using the latent indicator variable $Z_k$ we can construct a vector ${\bm{\theta}}_r \in \mathbb{R}^r: \{r \ll K\}$, from the elements of the weight vector ${\bm{\theta}}$ for which $Z_k=1$. With the reduced weight vector, ${\bm{\theta}}_r$ denoting the weight vector corresponding to the slab component, the DSS-prior is defined as \cite{mitchell1988bayesian,nayek2021spike},
\begin{equation}
    p\left( {{\bm{\theta }}|{\bm{Z}}} \right) = {p_{slab}}({\bm{\theta} _r})\prod\limits_{k,{Z_k} = 0} {{p_{spike}}({\bm{\theta} _k})},
\end{equation}
where ${p_{spike}}$ and ${p_{slab}}$ denotes the spike and slab distributions. Their distributions are defined as, ${p_{spike}}({\bm{\theta} _k}) = {\delta _0}$ and ${p_{slab}}({\bm{\theta} _r}) = \mathcal{N} \left( {{\bf{0}},{\sigma ^2}{\vartheta _s}{{\bf{R}}_{0,r}}} \right)$ with ${{\bf{R}}_{0,r}} = {{\bf{I}}_{r \times r}}$. Given the noise variance $\sigma^2$, the slab variance $\vartheta_s$ is assigned the Inverse-gamma prior with the hyperparameters ${\alpha _\vartheta }$ and ${\beta _\vartheta }$. The latent variables $Z_k$ take binary values; thus, the latent vector $\bm{Z}_k$ assigned the Bernoulli prior with common hyperparameter ${p_0}$. The hyperparameter ${p_0}$ is allowed to adapt according to the Beta prior with the hyperparameters ${\alpha _p }$, and ${\beta _p }$. The variance of the measurement noise is simulated from the Inverse-gamma distribution with the hyperparameters ${\alpha _\sigma }$, and ${\beta _\sigma }$. Taking into account these random variables, one can construct a Bayesian hierarchical model as presented in Fig. \ref{fig_graph}. In the model the hyperparameters ${\alpha _\vartheta }$, ${\beta _\vartheta }$, ${\alpha _p }$, ${\beta _p }$, ${\alpha _\sigma }$, and ${\beta _\sigma }$ are provided as a deterministic constants. For a review on the selection of hyperparameters, the readers are referred to Ref. \cite{o2009review}. The corresponding equations are given below.
\begin{subequations}
    \begin{align}
    p\left({\vartheta _s} \right) &= IG\left( {{\alpha _\vartheta },{\beta _\vartheta }} \right) \label{hyper1} \\ 
    p\left({Z_k}|{p_0} \right)  &= Bern\left( {{p_0}} \right);k = 1 \ldots K \label{hyper2} \\ 
    p\left( {p_0} \right) &= Beta\left( {{\alpha _p},{\beta _p}} \right) \label{hyper3} \\ 
    p\left({\sigma ^2} \right) &= IG\left( {{\alpha _\sigma },{\beta _\sigma }} \right). \label{hyper4}
\end{align}
\end{subequations}
    
From the DAG structure of the DSS-model in Fig. \ref{fig_graph}, the joint distribution of the random variables $p\left({\bm{\theta }}, {\bm{Z}}, {\vartheta _s}, {\sigma ^2}, {p_0}|{\bm{Y}}\right)$ can be expanded using the Bayes formula as,
\begin{equation}
\begin{aligned}
    p\left( {{\bm{\theta }},{\bm{Z}},{\vartheta _s},{\sigma ^2},{p_0}|{\bm{Y}}} \right)
    &= \dfrac{{p\left( {{\bm{Y}}|{\bm{\theta }},{\sigma ^2}} \right)p\left( {{\bm{\theta }}|{\bm{Z}},{\vartheta _s},{\sigma ^2}} \right)p\left( {{\bm{Z}}|{p_0}} \right)p\left( {{\vartheta _s}} \right)p\left( {{\sigma ^2}} \right)p\left( {{p_0}} \right)}}{{p\left( {\bm{Y}} \right)}}\\
    & \propto p\left( {{\bm{Y}}|{\bm{\theta }},{\sigma ^2}} \right)p\left( {{\bm{\theta }}|{\bm{Z}},{\vartheta _s},{\sigma ^2}} \right)p\left( {{\bm{Z}}|{p_0}} \right)p\left( {{\vartheta _s}} \right)p\left( {{\sigma ^2}} \right)p\left( {{p_0}} \right),
\end{aligned}
\end{equation}
where $p\left( {{\bm{Y}}|{\bm{\theta }},{\sigma ^2}} \right)$ denotes the likelihood function, $p\left( {{\bm{\theta }}|{\bm{Z}},{\vartheta _s},{\sigma ^2}} \right)$ is the prior distribution for the weight vector ${\bm{\theta}}$, $p\left( {{\bm{Z}}|{p_0}} \right)$ is the prior distribution for the latent vector ${\bm{Z}}$, $p\left( {{\vartheta _s}} \right)$ is the prior distribution for the slab variance ${\vartheta}_s$, $p\left( {{\sigma ^2}} \right)$ is the prior distribution for the noise variance, $p\left( {{p_0}} \right)$ is the prior distribution for the success probability $p_0$ and ${p\left( {\bm{Y}} \right)}$ is the marginal likelihood. Due to the DSS-prior, direct sampling of random variables from the joint distribution is intractable and thus requires Markov chain Monte Carlo (MCMC) algorithms. The Gibbs sampling technique is used in this framework to perform sampling from the joint distribution \cite{casella1992explaining}. For the Gibbs sampling, the conditional distributions of the random variables can be derived by referring to Ref. \cite{nayek2021spike}. The weights corresponding to the spike distribution do not contribute to the selection of the correct basis functions. Thus, only the weights corresponding to the latent variable $Z_k \ne 0$, i.e., the ${\bm{\theta}}_r$, are sampled using the Gibbs sampling. Referring to the Eqs.  \eqref{hyper1}, \eqref{hyper2}, \eqref{hyper3}, and \eqref{hyper4}, the sequence of the random variables ${{\bm{\theta }}^{(0)}},{\sigma ^{2(0)}},\vartheta _s^{(0)},p_0^{(0)},{{\bm{Z}}^{(0)}}, \ldots,{{\bm{\theta }}^{(1)}},{\sigma ^{2(1)}},\vartheta _s^{(1)},p_0^{(1)},{{\bm{Z}}^{(1)}}, \ldots $, using the Gibbs sampling technique is obtained by following steps,
\begin{enumerate}
    \item The weight vector ${\bm{\theta}}_r^{(i)}$ is sampled from the Gaussian distribution with mean ${{\bm{\mu}} _\theta }$ and variance ${{\bf{\Sigma}} _\theta }$ as,
    \begin{equation}\label{gibbs_theta}
        {{\bm{\theta }}_r^{(i)}}|{\bm{Y}},{\vartheta _s^{(i)}},{\sigma ^{2(i)}} \sim N\left( {{{\bm{\mu}} _\theta ^{(i)}},{{\bf{\Sigma}} _\theta ^{(i)}}} \right),
    \end{equation}
    where ${{\bm{\mu}} _\theta^{(i)} } = {\bf{\Sigma}}_\theta^{(i)} {\bf{L}}_r^{(i)T}{\bm{Y}}$ and ${{\bf{\Sigma}} _\theta ^{(i)}} = {\sigma ^{2(i)}}{\left( {{\bf{L}}_r^{(i)T}{{\bf{L}}_r^{(i)}} + \vartheta _s^{ {(i)}- 1}{\bf{R}}_{0,r}^{{(i)} - 1}} \right)^{ - 1}}$, respectively.
    
    \item The latent variable $Z_k^{(i+1)}$ is assigned the values from the set, $\left\{0,1 \right\}$ by using the Bernoulli distribution as,
    \begin{equation}\label{gibbs_z}
        {Z_k^{(i+1)}}|{\bm{Y}},{\vartheta _s^{(i)}},{p_0^{(i)}} \sim Bern\left( {{u_k}} \right),
    \end{equation}
    where ${u_k} = \frac{{{p_0}}}{{{p_0} + \lambda \left( {1 - {p_0}} \right)}}$ and $\lambda  = \frac{{p\left( {{\bm{Y}}|{Z_k^{(i)}} = 0,{{\bm{Z}}_{ - k}^{(i)}},{\vartheta _s^{(i)}}} \right)}}{{p\left( {{\bm{Y}}|{Z_k^{(i)}} = 1,{{\bm{Z}}_{ - k}^{(i)}},{\vartheta _s^{(i)}}} \right)}}$. Here, ${{\bm{Z}}_{ - k}^{(i)}} \in \mathbb{R}^{K-1}$ denotes the latent variable vector ${\bm{Z}}$ consisting of all the elements except the $k^{th}$ component. The probability that the $k^{th}$ latent variable $Z_k^{(i)}$ takes a value 0 or 1 is estimated as follows,
    \begin{equation}
    \begin{aligned}
      p\left( {{\bm{Y}}|{\bm{Z}}^{(i)},{\vartheta _s^{(i)}}} \right)
      &= \dfrac{{\Gamma \left( {{\alpha _\sigma } + \dfrac{N}{2}} \right)\beta _\sigma ^{{\alpha _\sigma }}}}{{\Gamma \left( {{\alpha _\sigma }} \right){{\left( {2\pi } \right)}^{\dfrac{N}{2}}}{{\left( {{\beta _\sigma } + \dfrac{1}{2}{{\bm{Y}}^T}{\bm{Y}}} \right)}^{\left( {{\alpha _\sigma } + \dfrac{N}{2}} \right)}}}}; \text{when all $\left\{ {Z_k^{(i)}}{: k = 1, \ldots ,K} \right\} = 0$} \\
     & = \dfrac{{\Gamma \left( {{\alpha _\sigma } + \dfrac{N}{2}} \right)\beta _\sigma ^{{\alpha _\sigma }}{{\left( {\left| {{\bf{R}}_{0,r}^{{(i)} - 1}} \right|\left| {{{\bf{\Sigma}} _\theta ^{(i)}}} \right|} \right)}^{\dfrac{1}{2}}}}}{{\Gamma \left( {{\alpha _\sigma }} \right){{\left( {2\pi } \right)}^{\dfrac{N}{2}}}\vartheta _s^{\dfrac{{{h_z}}}{2}}{{\left( {{\beta _\sigma } + \dfrac{1}{2}{{\bm{Y}}^T}{\bf{D}}{\bm{Y}}} \right)}^{\left( {{\alpha _\sigma } + \dfrac{N}{2}} \right)}}}}; \text{otherwise,}
        \end{aligned}
    \end{equation}
    where ${\bf{D}}=\left( {{{\bf{I}}_{N \times N}} - {{\bf{L}}_r^{(i)}}{{\bf{\Sigma}} _\theta ^{(i)}}{\bf{L}}_r^{(i)T}} \right)$.
    
    \item The noise variance ${\sigma ^{2(i+1)}}$ is simulated from the Inverse-gamma distribution as,
    \begin{equation}\label{gibbs_sigma}
        {\sigma ^{2(i+1)}}|{\bm{Y}},{\bm{Z}^{(i+1)}},{\vartheta _s^{(i)}} \sim IG\left( {{\alpha _\sigma } + \dfrac{N}{2},{\beta _\sigma } + \dfrac{1}{2}\left( {{{\bm{Y}}^T}{\bm{Y}} - {\bm{\mu}} _\theta^{(i)T} {\bf{\Sigma}} _\theta ^{{(i)} - 1}{{\bm{\mu}} _\theta ^{(i)}}} \right)} \right).
    \end{equation}
    
    \item The slab variance ${\vartheta _s^{(i+1)}}$ is sampled from the Inverse-gamma distribution as,
    \begin{equation}\label{gibbs_slab}
        {\vartheta _s^{(i+1)}}|{\bm{\theta ^{(i)}}},{\bm{Z}}^{(i+1)},{\sigma ^{2(i+1)}} \sim IG\left( {{\alpha _\vartheta } + \dfrac{{{h_z}}}{2},{\beta _\vartheta } + \dfrac{1}{{2{\sigma ^2}}}{\bm{\theta }}_r^{(i)T}{\bf{R}}_{0,r}^{{(i)} - 1}{{\bm{\theta }} _r^{(i)}}} \right).
    \end{equation}
    
    \item The success rate ${p_0^{(i+1)}}$ is sample from the Beta distribution as,
    \begin{equation}\label{gibbs_p0}
        {p_0^{(i+1)}}|{\bm{Z}}^{(i+1)} \sim Beta\left( {{\alpha _p} + {h_z},{\beta _p} + K - {h_z}} \right),
    \end{equation}
    where ${h_z} = \sum\nolimits_{k = 1}^K {{Z_k}^{(i+1)}}$.
    
    \item The weight vector ${\bm{\theta}}_r^{(i+1)}$ is updated using the step 1.
\end{enumerate}
In the MCMC iterations, the initial 500 samples are discarded as burn-in samples. Let ${N_s}$ denote the number of MCMC required to achieve the stationary distribution after burn-in samples are discarded. The probability of choosing an arbitrary function in the final model can be estimated from the marginal posterior inclusion probability (PIP) criterion. The PIP:= $p\left( {{Z_k} = 1|{\bm{Y}}} \right); \forall 1, \ldots, K$ can be estimated by taking the mean over the Gibbs samples for each of the $k^{th}$ latent vector ${Z_k}$ \cite{nayek2021spike} as
\begin{equation}\label{mpip}
    p\left( {{Z_k} = 1|{\bm{Y}}} \right) \approx \dfrac{1}{{{N_s}}}\sum\limits_{j = 1}^{{N_s}} {Z_k^{(j)}} ;{\rm{ }}k = 1, \ldots ,K.
\end{equation}
The basis functions with $p\left( {{Z_k} = 1|{\bm{Y}}} \right) > 0.5$ are included in the final model. A PIP value greater than 0.5 indicates that the corresponding basis functions are observed in at least 50\% MCMC simulations. Let there be $r$ ($r<<K$) basis functions in the final model. Once the correct basis functions are selected based on the PIP$>$0.5, the expected value of the weights can be found from the posterior distributions of the corresponding weights as
\begin{equation}
    {\mathbb{E}}\left[ \bm{\theta}_k \right] \approx \dfrac{1}{{{N_s}}}\sum\limits_{j = 1}^{{N_s}} {{\bm{\theta}}_k^{(j)}} ;{\rm{ }}k = 1, \ldots ,r.
\end{equation}
The above operation will produce the weight vector $\bm{\theta} \in \mathbb{R}^{K}$ that will have non-zero values at those indices where $p\left( {{Z_k} = 1|{\bm{Y}}} \right) > 0.5$. For a $m$-dimensional dynamical system driven by $n$ unknown forces, it will result in ${K \times m}$ dimensional drift weight matrix $\bm{\theta}^f \in \mathbb{R}^{K \times m}$, where each column corresponds to a DOF. The weight matrix corresponding to the covariance of the diffusion will have a dimension $\bm{\theta}^g \in \mathbb{R}^{K \times (m \times m)}$ with $m^2$ number of columns, each column is associated with an element in the covariance matrix. Given, the present state vector $\bm{X}_j \in \mathbb{R}^{1 \times m}$, one can construct the library ${\bf{L}} \in \mathbb{R}^{K}$. The drift vector thus can be obtained as, ${\hat{\bm{f}}\left(\bm{X}_t, t \right) \in \mathbb{R}^{m}} = {\bf{L}}{\bm{\theta^f}}$. In the case of the diffusion component, the covariance matrix is obtained as ${\hat{\bf{\Gamma}}\left(\bm{X}_t, t \right)} \in \mathbb{R}^{m \times m} = \mathcal{T}\left( {\bf{L}}{\bm{\theta^g}} \right)$, where $\mathcal{T}: \mathbb{R}^{1 \times (m \times m)}$ is a reshaping transformation that transforms the output ${\bf{L}}{\bm{\theta^g}} \in \mathbb{R}^{1 \times (m \times m)}$ into a matrix of size $m \times m$. one can find the evolution of the states using following relations,
\begin{equation}\label{internal}
    \bm{X}_{j+1} = \bm{X}_{j} + \hat{\bm{f}}\left(\bm{X}_t,\bm{u}_t, t \right)^T \Delta t + {\sqrt{ \hat{\bf{\Gamma}}\left(\bm{X}_t,\bm{u}_t, t \right)}} \Delta \bm{B}_t.
\end{equation}
The steps involved in the governing model discovery step are illustrated in Algorithm \ref{algvec1}.
\begin{algorithm}
	\caption{Discovery of governing model}\label{algvec1}
	\begin{algorithmic}[1]
	\Require{Get output measurements: ${\bf{X}}(t) \in \mathbb{R}^{N \times m}$, initialize: $\{ \alpha_p, \beta_p, \alpha_\sigma, \beta_\sigma, \alpha_\vartheta, \beta_\vartheta, p_0^{(0)}, \vartheta_s^{(0)} \}$.}
	    \State{Obtain the library ${\bf{L}}_{\mu} = {\mathbb{E}}[{\bf{L}}]$.}
	    \For {$i = 1, \ldots, m$ ($i^{th}$ diffusion process)}
	    \State{Estimate: ${\bm{f}_i}\left( {{{\bm {X}}_t},{\bm{u}_t},t} \right)$, ${{\bf{\Gamma}}_{ij}}\left( {{{\bm {X}}_t},{\bm{u}_t},t} \right)$.} \Comment{Eq. \eqref{kramers}}
	    \State{Obtain the target vectors ${{\bm{Y}}_i}$ and ${{\bm{Y}}_{ij}}$.}  \Comment{Eq. \eqref{regress}}
	    \State{Estimate: $\sigma^{2,(0)}$ = Var(${{\bf{L}}_{\mu}}{\bm{\theta}}-{\bm{Y}}$).}
	    \State{Find: $\bm{Z}^{(0)}$=$\left[Z_1^{(0)}, \ldots, Z_K^{(0)} \right]$ $\ni$ MSE$({\bf{L}}_{\mu}{\bm{\theta}}-{\bm{Y}})$ is minimum.}
	    \State{Find: ${\bm{\mu}}_\theta$, ${\bf{\Sigma}}_\theta$ and ${\bm{\theta}}^{(0)}_{i}$.} \Comment{Eq. \eqref{gibbs_theta}}
		\For {$l$ = $1, \ldots, {N_{MCMC}}$}	
		\State Update ${\bm{Z}}^{(l)}, \sigma^{2(l)}, \vartheta _s^{(l)}, p_0^{(l)}$ and ${\bm{\theta}}^{(l)}_{i}$. \label{MCMC1} \Comment{Eq. \eqref{gibbs_z} - Eq. \eqref{gibbs_theta}}
		\EndFor
		\State{Discard the burn-in MCMC samples}
		\For {$k = 1, \ldots, K$}
    	\State{Estimate PIP values: $p(Z_k=1|{\bm{Y}})$} \Comment{Eq. \eqref{mpip}}
    	\State{Select basis functions $\ni$ $p(Z_k=1|{\bm{Y}}) > 0.5$.}
    	\State{Set ${\bm{\theta}_{i_{k}}}$ = 0; if $p(Z_k=1|{\bm{Y}}) \ngtr 0.5$.}
    	\State{Find: ${\mathbb{E}}[{\bm{\theta}_{i_{k}}}]$ and ${\sigma}[{\bm{\theta}_{i_{k}}}]$.}
		\EndFor
	    \EndFor
	\Ensure{$\{ {\bm{\theta}}_{i};i=1 \ldots {m} \}$, $\{ \vartheta_i ({\bm{X}}); i=1 \ldots {m} \}$.}
	\end{algorithmic}
\end{algorithm}

\begin{figure*}
\begin{center}
\includegraphics[width=\textwidth]{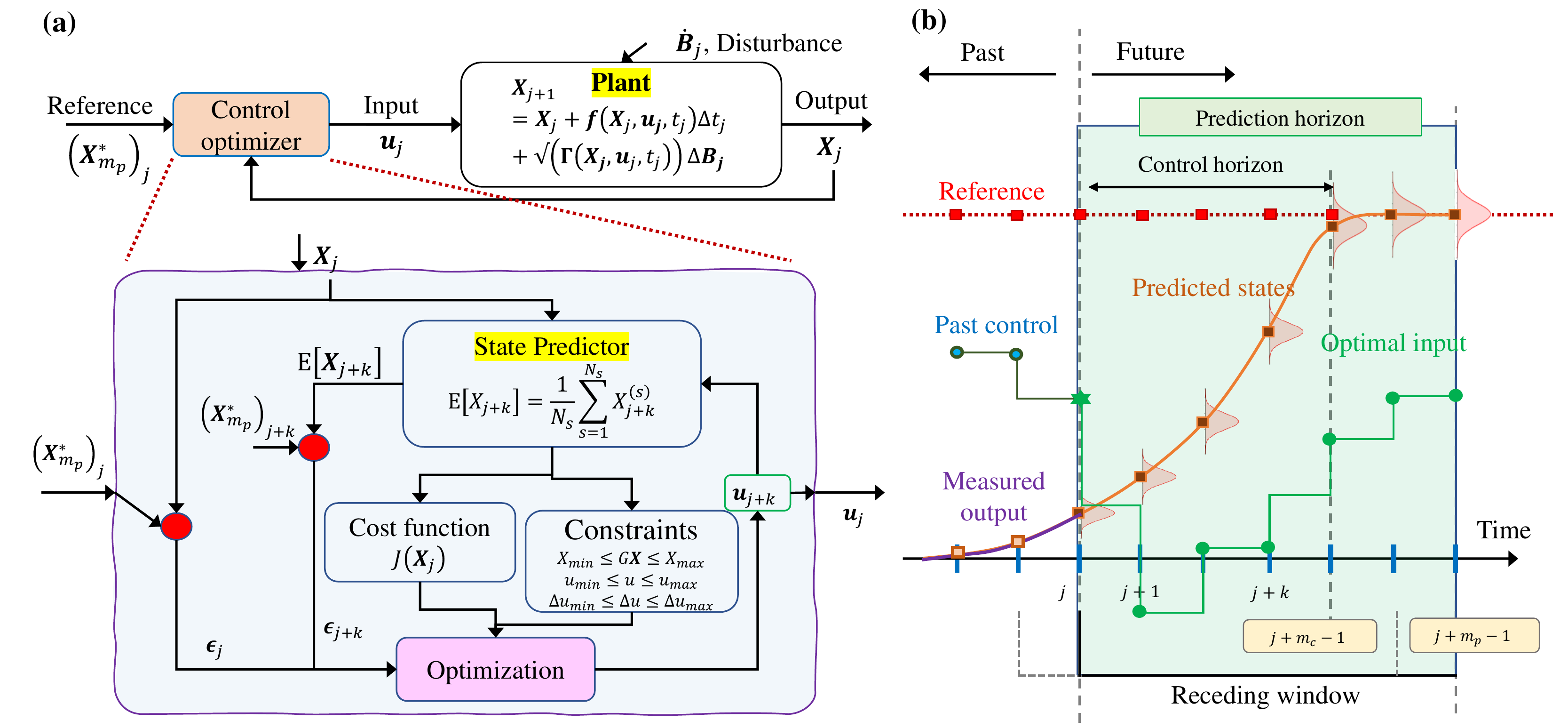}  
\caption{Schematic of the stochastic model predictive control. (a) The control loop. (b) The receding window. Given the reference trajectory, the recent measurements of the state trajectory, and the control force, the MPC obtains the expectation of the predictions. Based on the expected predictions of the state, the control force is optimized over the control horizon, and the first in the sequence is provided as input.} 
\label{fig_MPC}         
\end{center}   
\end{figure*}
\subsection{Stochastic model predictive control (MPC)}
MPC is an iterative optimal control problem that solves a finite horizon open-loop optimal control problem at each instant of the fine time horizon. The finite time horizon defines the time window over which the prediction is done using the identified model. As a result of the optimization problem, the sequence of control inputs over the time horizon is obtained from which only the first control variable is used as input. This operation is performed repeatedly by sliding the horizon window, thus sometimes referred to as moving horizon control or receding horizon control, yielding a new control and a new predicted state path. The choice of the frequency of the control depends on whether the control required is continuous or intermittent. MPC is a powerful tool since it explicitly learns the underlying governing physics of the process and thus can keep track of the abrupt changes due to external disturbances. However, when the input force in the process is not known in advance, the behavior of the system in the future can not be predicted correctly. In such cases, we can harness the power of Stochastic Model Predictive Control (SMPC) \cite{hokayem2012stochastic,dai2015cooperative,ning2021online}. SMPC takes into account the external disturbances in the design process and allows to the design of controllers which optimize the expected value of the controllable variable rather than the worst case. The essence of the SMPC is illustrated in Fig. \ref{fig_MPC}.

\begin{algorithm}
	\caption{MPC framework}\label{algvec2}
	\begin{algorithmic}[1]
	\Require{Get output measurements: ${\bf{X}}(t) \in \mathbb{R}^{N \times m}$, initialize hyperparameters, define the control parameters: $m_c$, $m_p$, $\tilde{\Delta t}$, ${\bf{R}}_{\bm{u}}$, ${\bf{R}}_{\Delta {\bm{u}}}$, ${\bf{Q}}_{m_p}$, ${\bf{Q}}$, $\bm{X}^*_{m_p}$, $\bm{X}(0)$, $\bm{u}(0)$.}
	    \State{Define the bounds on the constraints.}
	    \For {$j = 1, \ldots, T$ (time horizon)}
	    \State{Update governing model using Algorithm \ref{algvec1}.}
	    \For {$k = 1, \ldots, m_p$ (prediction horizon)}
	    \State{Generate the predictions: $\{ \bm{X}_{j}^{(s)}; \; s=1, \ldots, {N_c} \}$.}
	    \State{Find: ${\mathbb{E}}[\bm{X}_j] \approx \frac{1}{{{N_c}}}\sum_{s = 1}^{{N_c}} {{\bm{X}_j}^{(s)}}$.}
	    \State{Find: ${\mathbb{E}}[{\mathcal{G}}({\bm{X}_j})] \approx \frac{1}{{{N_c}}}\sum_{s = 1}^{{N_c}} {{\mathcal{G}}({\bm{X}_j})^{(s)}}$.}
	    \State{Find the expected value of constraints.}  \Comment{Eq. \eqref{bounds}}
	    \State{Find the cost function: $J({\bm{X}_j})$.} \Comment{Eq. \eqref{costfun}}
	    \EndFor
	    \State{Obtain ${\bm{u}}(.| \bm{X}_{j+k}); \; k= 1, \ldots, {m_p}$.}
	    \State{Use first value as input for next model update.}
	    \EndFor
	\Ensure{Controlled response: $\{ {\bm{X}^{*}_{j}}; \; j=1, \ldots, T \}$ }.
	\end{algorithmic}
\end{algorithm}

The central idea is to obtain the optimal sequence of the control inputs ${\bm{u}}(.| \bm{X}_j) := \left\{\bm{u}_{j+k}; \; k=1, \ldots, m_c \right\}$ over the control horizon $T_c = {m_c}{\tilde{\Delta t}}$ by optimizing an appropriate cost function $\bm{J}(\bm{X}_j)$ over the prediction horizon $T_p = {m_p}{\tilde{\Delta t}}$; with ${\tilde{\Delta t}}$ being the time step of the controller. In general, ${\tilde{\Delta t}}$ and the sampling rate can be both identical and/or different; in the identical case, they will mean that the control is being done actively at each instant of time. The terms $m_c$ and $m_p$ refer to the control and prediction horizon such that ${T_c} \le {T_p}$. During the optimization, the prediction of the process states $\bm{X}_{j+k}; \; k=1, \ldots, m_p$ are obtained using Eq. \eqref{internal}. In the sequence ${\bm{u}}(.| \bm{X}_j)$, only the first control value $\bm{u}_{j+1}$ is used as input in the system. This process is repeated with $\bm{u}_{j+1}$ as the initial value for the next instant in a loop for all the time instances to obtain the optimal control sequence ${\bm{u}}(.| \bm{X}_j); \; j= 1, \ldots, T$. To summarize, the following feedback control law can be referred,

\begin{equation}\label{feedback}
    \bm{u}\left( {j+1} | {\bm{X}_j} \right) = {\bm{u}}_{j+1}.
\end{equation}
We consider a quadratic function of the state and the input as the performance index and treat the expected value of the performance index as the cost function. The cost function $\bm{J}(\cdot)$ over the receding horizon to be optimized is given as \cite{di2013stochastic,ning2021online},
\begin{equation}\label{costfun}
    \begin{aligned}
    \mathop {\min }\limits_{u(.|{\bm{X}_j})} J({\bm{X}_j}) =& \mathop {\min }\limits_{u(.|{\bm{X}_j})} {\mathbb{E}} \Biggl[ \left\| {{\bm{X}_{j + {m_p}}} - \bm{X}_{{m_p}}^*} \right\|_{{{\mathbf{Q}}_{{m_p}}}}^2 + \sum\limits_{k = 0}^{{m_p} - 1} {\left\| {{\bm{X}_{j + k}} - \bm{X}_{{m_p}}^*} \right\|_{\mathbf{Q}}^2}  + \\
    &\sum\limits_{k = 1}^{{m_c} - 1} { \left( \left\| {{\bm{u}_{j + k}}} \right\|_{{\mathbf{R}_{\bm{u}}}}^2 + \left\| {\Delta {\bm{u}_{j + k}}} \right\|_{{\mathbf{R}_{\Delta \bm{u}}}}^2 \right)} \Biggr] \quad \forall j = 1, \ldots, T,
    \end{aligned}
\end{equation}
subjected to the internal process dynamics in Eq. \eqref{internal}. Here, $\bm{X}_{j+k}; \; k=1, \ldots, m_p$ is the $k^{th}$ vector of controlled variable, $\bm{X}_{m_p}^*$ is the vector of a reference variable, ${\bm{u}}_{j+k}; \; k=1, \ldots, m_p$ is the $k^{th}$ control variable, ${\bf{Q}} \ge 0$ and ${\bf{Q}}_{m_p} \ge 0$ are the weight matrices defining the relative importance among the process states, ${\mathbf{R}}_{\bm{u}} > 0$ and ${\mathbf{R}}_{\Delta \bm{u}} > 0$ are the weight matrices for the control variable ${\bm{u}}_{j+k}$ and the control rate $\Delta {\bm{u}}_{j+k} = {\bm{u}_{k+1}} - {\bm{u}_{k}}$, respectively. The weight matrices ${\bf{Q}}$ and ${\bf{Q}}_{m_p}$ penalizes the deviations of the predicted state $\bm{X}_{k}$ from the reference value, the weight ${\mathbf{R}}_{\bm{u}}$ prevents the control variables from taking very large values and the weight ${\mathbf{R}}_{\Delta \bm{u}}$ penalizes the large relative changes in the control inputs. The notation $\left\|(.) \right\|_{\bf{A}}^{2}$ denotes the matrix norm of a vector. The optimization is subjected to the following equality and inequality constraints on the process state and control inputs,
\begin{subequations}\label{bounds}
    \begin{align}
        & \mathcal{X}^l_{\min} \le {\bf{A}}{\mathbb{E}}[{\bm{X}_j}] \le \mathcal{X}^{l}_{\max} \\
        & \mathcal{X}^{nl}_{\min} \le {\mathbb{E}}[\mathcal{G}({\bm{X}_j})] \le \mathcal{X}^{nl}_{\max} \\
        & {\bm{u}_{\min }} \le {\bm{u}_k} \le {\bm{u}_{\max }}\\
        & \Delta {\bm{u}_{\min }} \le \Delta {\bm{u}_k} \le \Delta {\bm{u}_{\max }},
    \end{align}
\end{subequations}
where $\mathcal{G}(\bm{X})$ denotes some arbitrary nonlinear function of process states. The pseudo-code of the SMPC in the proposed MASMPC framework is demonstrated in Algorithm \ref{algvec2}.

\section{Numerical experiments}\label{sec:result}
For the purpose of validation, the proposed MASMPC approach is implemented on three numerical examples in this section. First, a predator-prey model is considered, where the interaction between the two species is stabilized. Then, the Lorenz oscillator is considered to validate the applicability of the proposed approach for chaotic systems. Finally, a semi-active vibration control device using the active tuning of the damping parameter of the tuned mass damper (TMD) using the proposed control strategy is demonstrated on a real 76-storey benchmark structure. Broadly, for each of the problems, it is assumed that no knowledge of the input force in the system is available. With this restriction, first, the physics of the underlying dynamical systems is identified using the Bayesian System Identification of Stochastic Systems (BSISS) algorithm. Then an ensemble of predicted system responses is generated via stochastic simulation. For this purpose, Euler-Maruyama (EM) stochastic integration scheme is employed \cite{kloeden1992higher}. From these ensembles, the expected objective cost function, along with the constraints, is obtained.

For model discovery, the hyperparameters are set as: $a_p$=0.1, $b_p$=1, $a_v$=0.5, $b_v$=0.5, $a_\sigma$=$10^4$, and $b_\sigma$=$10^4$ \cite{o2009review}. The random parameters in the hierarchical model are initialized as $p_0^{(0)}$=0.1, $\vartheta^{(0)}$=10, and $\sigma ^{2(0)}$ is set equal to the residual variance from ordinary least-squares regression. The initial binary latent vector is obtained by setting ${\bm{Z}}^{(0)}=[Z_1, \ldots, Z_K]$ to zero and then activating the components of ${\bm{Z}}$ such that the mean-squared error MSE$({\bf{L}}_{\mu}{\bm{\theta}}-{\bm{Y}})$ in the ordinary least-squares is minimum. Given all the other parameters, the initial value of ${\bm{\theta}}^{(0)}$ is obtained from Eq. \eqref{gibbs_theta}. Markov chains of length 2000 with 500 burn-in sizes is used for the model identification. For all the demonstrations, the data are simulated using the EM scheme with the parameters listed in Table \ref{table_param}. The noise in the signals is empirically modeled as $N$-dimensional sequence of zero-mean Gaussian white noise whose intensity is set equal to 5\% of the standard deviation of the signal. The dictionary ${\bf{L}} \in \mathbb{R}^{N \times K}$ is constructed from 5 different types of mathematical functions \cite{nayek2021spike}, an identity function ${\bf{1}} \in \mathbb{R}^{N}$, the polynomial functions ${P^\mathcal{P}}({\bm{X}}) \in \mathbb{R}^{N \times m}$ with highest degree 6, the signum function $sgn({\bm{X}}) \in \mathbb{R}^{N \times m}$, the modulus function ${\left| {\bm{X}} \right|} \in \mathbb{R}^{N \times m}$, the tensor product ${{\bm{X}} \otimes \left| {\bm{X}} \right|} \in \mathbb{R}^{N \times {2m}}$, each function representing a mapping of the $m$-dimensional state vector $\bm{X}$.

In this work, the control is achieved in two stages, firstly, the governing physics is identified from the sensor data, and then the optimal control force is obtained using the proposed MASMPC framework. The identification of the governing physics is performed from 200 ensembles of 1s data obtained at a sampling frequency of 1000Hz. 
The performances of the proposed framework are compared with control results performed on predictions made from long short-term memory (LSTM) networks \cite{zhang2019deep}. For training the LSTM network, the same ensemble of 1s data is utilized, and a sliding window of length 20 with 10 hidden states is considered. The Adam optimizer with a learning rate of 0.001 and a total of 1000 epochs is considered to train the network.

\begin{table*}[ht]
	\centering
	\caption{Simulation parameters of the systems. }
	\begin{tabular}{lll}
		\hline
		\textbf{Systems} & \textbf{Simulation parameters} & \\ \cline{2-3}
		& \textbf{System} & \textbf{Control} \\ 
		\hline
		Lotka-Volterra$^{\ref{example1}}$ & $a$=0.5, $b$=0.025, $c$=0.5, $d$=0.005 & $Q$=diag(1, 1), $R$=0.5, $R_u$=0.5 \\
		& $\sigma_1$=0.2, $\sigma_2$=0.2, $[{\bm{X}}(0)]$ = $[60, 50]$ & $m_c$=10, $m_p$=10, $N_c$=100 \\
		\hline
		Lorenz$^{\ref{example2}}$ & $\alpha$=10, $\rho$=28, $\beta$=8/3, $\sigma_1$=4, $\sigma_2$=4 & $Q$=diag(1, 1, 1), $R$=0.001, $R_u$=0.001 \\
		& $\sigma_3$=4, $[{\bm{X}}(0)]$ = $[-8, 8, 27]$ & $m_c$=10, $m_p$=10, $N_c$=200 \\
		\hline
		SATMD$^{\ref{sec:example3}}$ & Refer \cite{yang2004benchmark} & $Q$=diag(1, 1, 0, 0), $R$=0.1, $R_u$=0.1 \\
		& & $m_c$=10, $m_p$=10, $N_c$=100 \\
		\hline
	\end{tabular}
	\label{table_param}
\end{table*}

\subsection{Population dynamics: Lotka-Volterra model}\label{example1}
Predator-prey models are widely used for modeling the evolution of two species in a competitive environment \cite{brauer2012mathematical}. One of the simplifications over the general Kolmogorov's predator-prey model is the Lotka-Volterra Model. Beyond the predator-prey interactions, the Lotka-Volterra model is also used for modeling the relations between resource-consumer, plant-herbivore, parasite-host, tumor cells (virus)-immune system, susceptible-infectious interactions, etc. Due to its wide application in various fields, the proposed scheme is implemented on the Lotka-Volterra Model as the first case study. The solution of the model forms a periodic oscillator. If it is assumed that the two species do not go extinct with time, then the periodic solution has a unique equilibrium. The aim of this study is to obtain the corresponding external enforcement required to stabilize the interactions of the two species. In its original format, the Lotka-Volterra model represents the interactions between the two species using a deterministic differential equation model; however, the interactions are not always deterministic since the uncertainty in their mathematical model always exists. We leverage the stochastic theory and represent the uncertainty in the population growth of the species by Brownian motions \cite{kloeden1992higher}. The governing stochastic dynamics of the Lotka-Votera model subjected to Brownian motion with control is given by the following equation,
\begin{equation}\label{eq_lotka}
    \begin{aligned}
    &{\dot{X}_1} = a{X_1} -b{X_1}{X_2} + \sigma_1 \dot{B}_1 \\
    &{\dot{X}_2} = -c{X_2} +d{X_1}{X_2} + \sigma_2 \dot{B}_2 + u,
    \end{aligned}
\end{equation}
where $\{X_1, X_2 \}$ are the states of the model, $\{a, b, c, d\}$ are deterministic parameters describing the growth rate of prey, impact of prey population on the predator, the death rate of the predator, and rate of growth of the predator population. The terms $\{\dot{B}_1, \dot{B}_2 \}$ denote the Brownian motion characterizing the uncertainty in the population prediction, and the terms $\{\sigma_1, \sigma_2 \}$ represents the degree of uncertainty. The term $u$ is responsible for the control. The above set of differential equations can be converted into equivalent SDEs of the form Eq. \eqref{sdeg}.

\begin{figure}[ht!]
     \centering
     \begin{subfigure}[b]{0.49\textwidth}
         \centering
         \includegraphics[width=\textwidth]{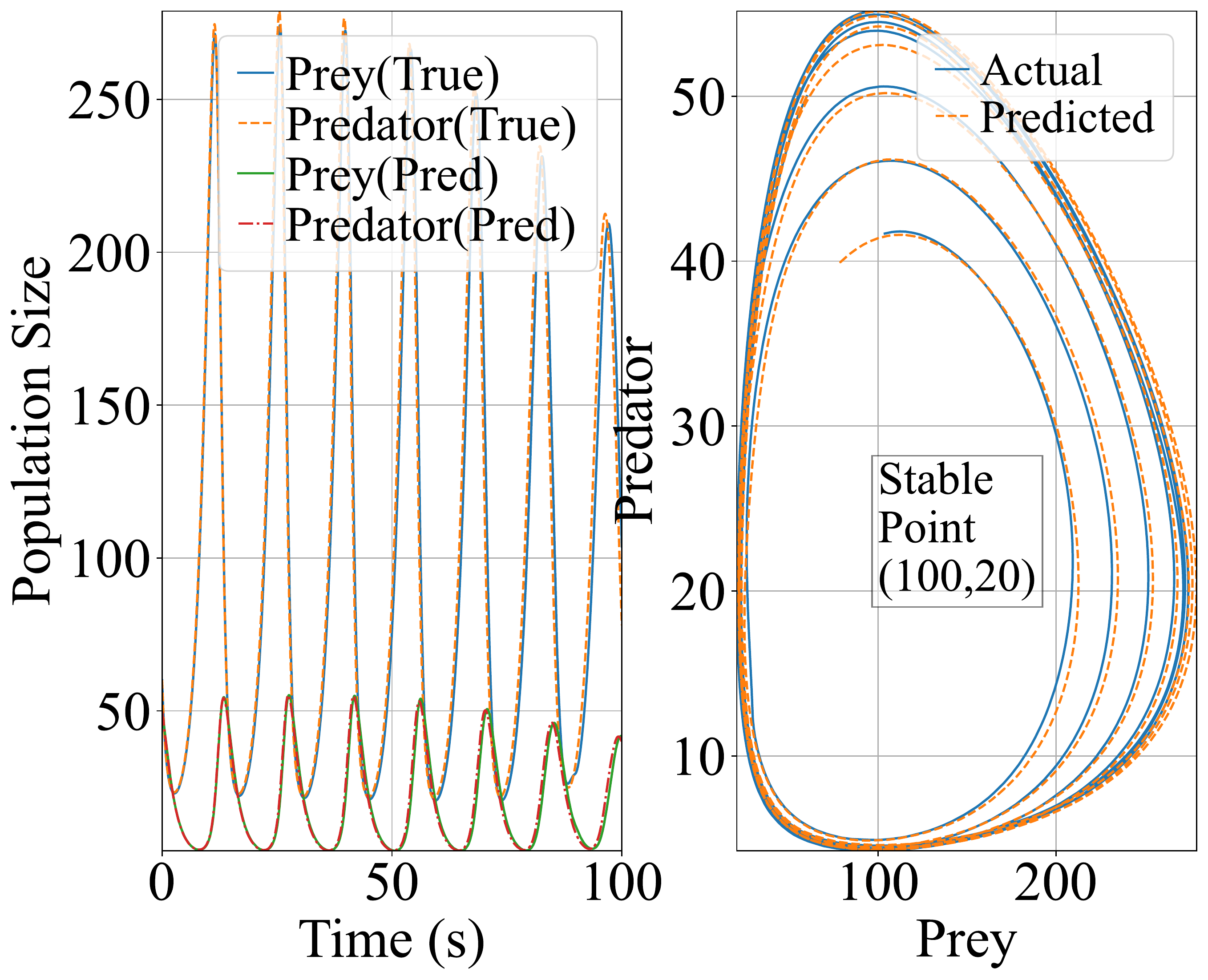} 
        \caption{Prediction}
        \label{fig_lotka1} 
     \end{subfigure}
     \hfill
     \begin{subfigure}[b]{0.5\textwidth}
         \centering
         \includegraphics[width=\textwidth]{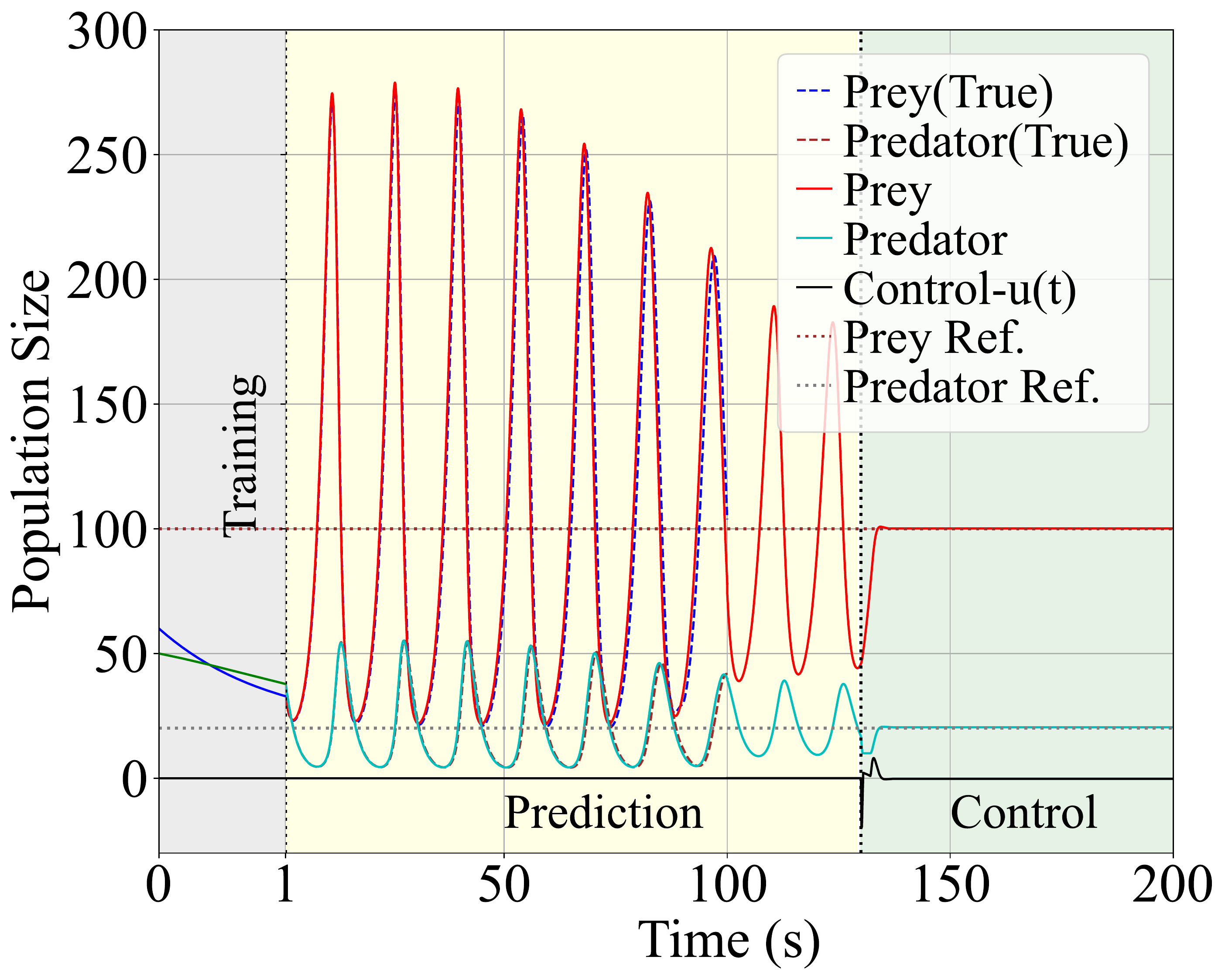} 
        \caption{Control}
        \label{fig_lotka2}  
     \end{subfigure}
     \caption{Validation and control performance for the Lotka-Volterra model: the training is performed using one-second data, after which the model is validated for 100 seconds. Then the long-term prediction is continued for 30 seconds. Finally, the control is turned on at 130$^{th}$ second. Note that the prediction phase is not required for doing control and is presented here only for validation purposes. The control phase can start immediately after the training phase (i.e., 1s onward).}
     \label{fig:lotka_gen}
\end{figure}

The system is simulated using the parameters given in Table \ref{table_param}. The steady-state solution of the population is given by $\bm{X}^{\rm{crit}} = \{a/b, c/d \}=\{100, 20 \}$. Note that we assume that the underlying physics is not known a priori; instead, we only have access to noisy data. The objective is to stabilize the population dynamics at this point. The control is carried out at a rate of $\tilde{\Delta t}$=0.1 second. In addition to the control parameters in Table \ref{table_param}, following inequality constraints are enforced, (i) $X_2 > 10$ and (ii) -$20< u <20$.

Figure \ref{fig_lotka1} shows the prediction upto 100s using the learned model. It can be seen that the predicted response (mean) using the proposed framework almost accurately emulates the original response. This allows the model to adjust for system disturbances and hard changes that the model was not trained for. In Fig. \ref{fig_lotka2}, the complete process of identification, prediction, and control is presented. First, one-second data is used for training, followed by verification for up to 100 seconds, long-term prediction for 30 seconds, and finally, the control is turned on. Once the control is turned on, the proposed framework quickly adjusts the system to the target value, as observed in Fig. \ref{fig_lotka2}. This exhibits the stability of the proposed framework, which can discover the correct physics, predict the correct future, and offer control feedback to the system without requiring input force information. The performance of the proposed framework is compared with modern deep learning-based LSTM networks, the results of which are provided in Table. \ref{tab:efficiency}. The absolute prediction error is obtained with respect to the original predictions made using EM scheme, whereas, in the control stage, the errors are computed by measuring the distance between the reference and control stage. It is evident that both the predictions and the controlled responses obtained using the proposed framework are nearer to the original predictions and reference points as compared to the results obtained using LSTM.

\begin{table}[ht!]
    \centering
    \begin{tabular}{p{4cm}p{2cm}llp{0.5cm}ll}
        \hline
        \multirow{3}{*}{Systems} && \multicolumn{5}{c}{Absolute mean error} \\ \cline{3-7}
        && \multicolumn{2}{c}{MASMPC} && \multicolumn{2}{c}{LSTM} \\ \cline{3-4} \cline{6-7}
        && Prediction & Control && Prediction & Control \\
        \hline
        Lotka-Volterra$^{\ref{example1}}$ & Prey & 0.0964 & 0.0011 && 0.7066 & 0.5489 \\
        & Predator & 0.0894 & 0.0201 && 1.1384 & 1.2728 \\ \cline{1-7}
        Lorenz$^{\ref{example2}}$ & X & 8.4222 & 0.0219 && 154.6601 & 11.6147 \\
        & Y & 5.6693 & 0.0164 && 290.3337 & 14.6616 \\
        & Z & 0.0181 & 0.0102 && 0.098881 & 0.7981 \\
        \hline
    \end{tabular}
    \caption{Efficiency of proposed MASMPC framework against LSTM networks}
    \label{tab:efficiency}
\end{table}

\begin{figure}[ht!]
     \centering
     \begin{subfigure}[b]{\textwidth}
         \centering
         \includegraphics[width=\textwidth]{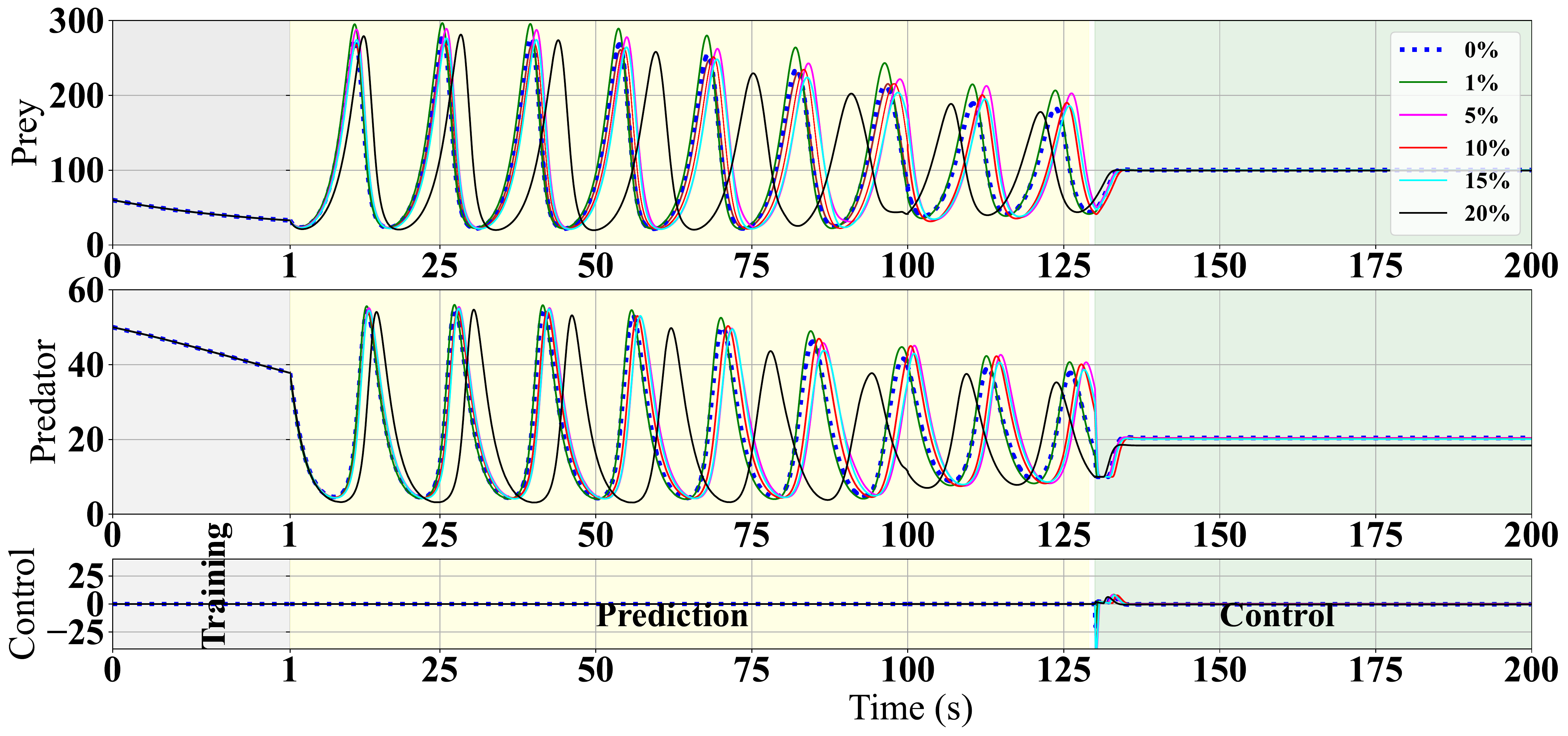} 
        \caption{Time series in validation and controlled stage for increasing measurement noise}
        \label{fig_lotka_noise}  
     \end{subfigure}
     \hfill
     \begin{subfigure}[b]{0.48\textwidth}
         \centering
         \includegraphics[width=\textwidth]{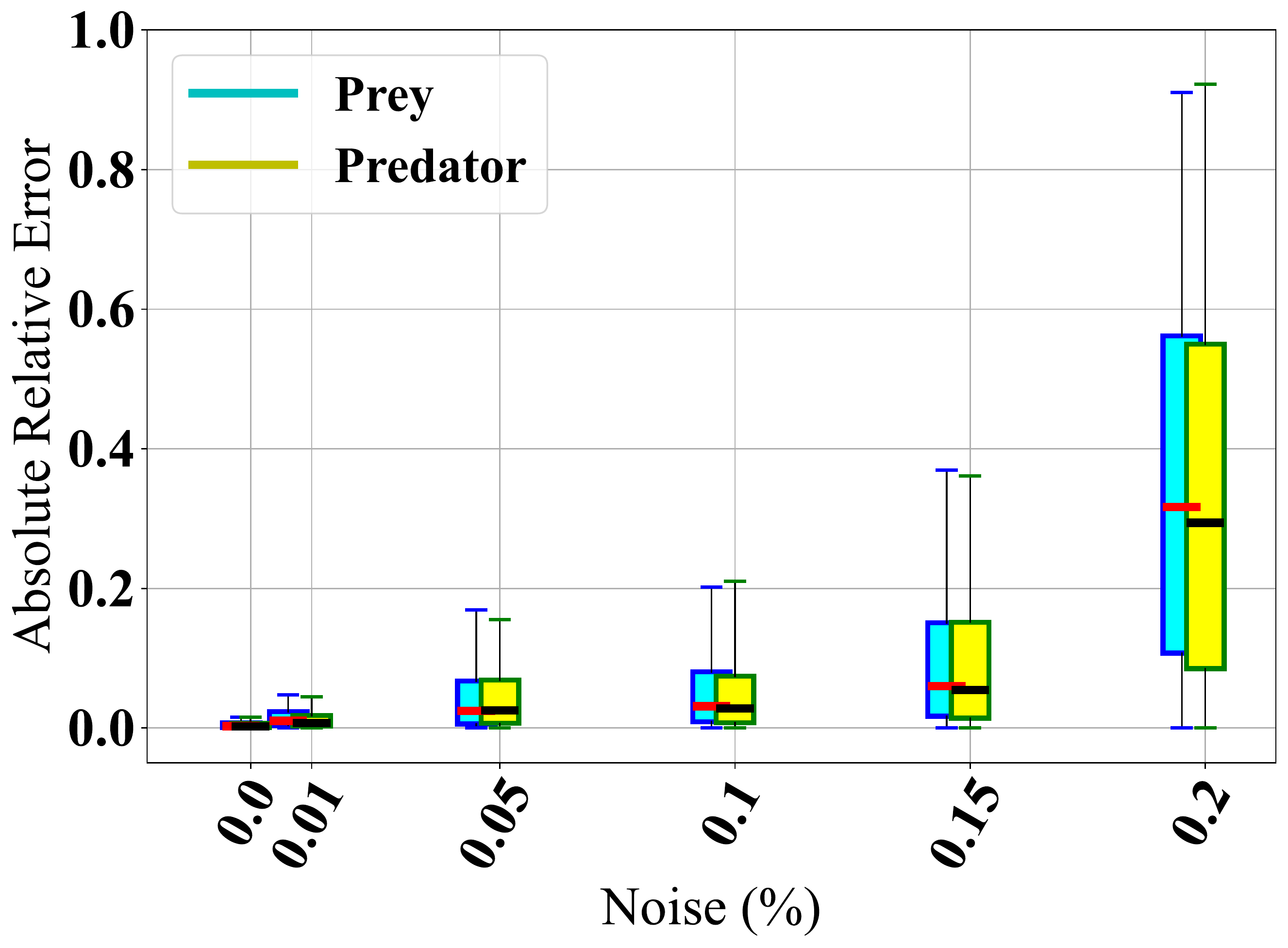} 
        \caption{Error in the validation set for Prey}
        \label{fig_lotka_error_val} 
     \end{subfigure}
     \hfill
     \begin{subfigure}[b]{0.48\textwidth}
         \centering
         \includegraphics[width=\textwidth]{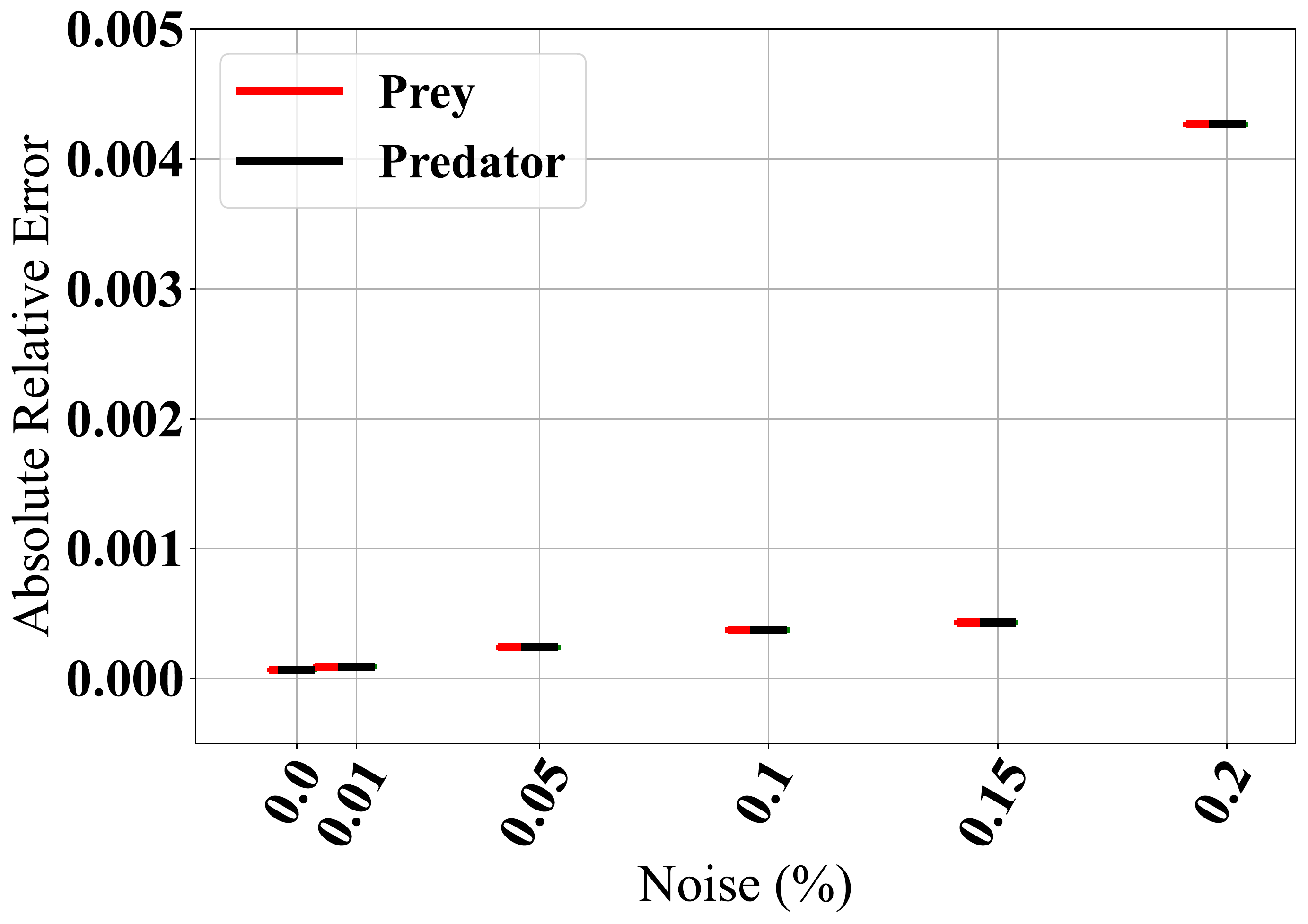} 
        \caption{Error in the validation set for Predator}
        \label{fig_lotka_error_cont} 
     \end{subfigure}
        \caption{Prediction and control performance for increasing measurement noise: the performance of the proposed framework is tested on 5 levels of measurement noise (1\%, 5\%, 10\%, 15\%, 20\%). The absolute relative error is obtained in terms of the noiseless response. In (b), it is evident the proposed framework is able to predict the correct response even when the input is corrupted with a noise of magnitude upto 15\% of the standard deviation of input. The (c) conveys that even if the prediction becomes faulty, the proposed framework is able to correct the response without any greater difficulty.}
        \label{fig:lotka_case_i}
\end{figure}

\subsubsection{Sensitivity to noise}
The presence of noise in the sensor measurements is a practical issue and the soundness of an efficient algorithm to such measurement noises is among the primary requirements for industrial implementations. Here, we test the robustness of the proposed framework with noise intensities ranging from 1\% to 20\% of the standard deviation of the output signal. The results are presented in Fig. \ref{fig_lotka_noise}. The absolute relative error is estimated for both the validation and controlled stage and depicted along with their statistics in Fig. \ref{fig_lotka_error_val} and \ref{fig_lotka_error_cont}. From the values of the errors in the figures, it is understood that the proposed MASMPC algorithm (i) is able to effectively discover the governing stochastic model from measurements corrupted with noises of intensities upto 15\% without requiring the need for input measurements and (ii) is able to drive the discovered model to given reference points. Further, it is quite clear from Fig. \ref{fig_lotka_error_cont} that even when the performance of the discovery module gets suffered from noises of the order of more than 15\%, the control module is able to compensate for the discrepancy in the governing physics and correctly control the intended system as soon as the control is switched on.

\begin{figure}
     \centering
     \begin{subfigure}[b]{0.5\textwidth}
         \centering
         \includegraphics[width=\textwidth]{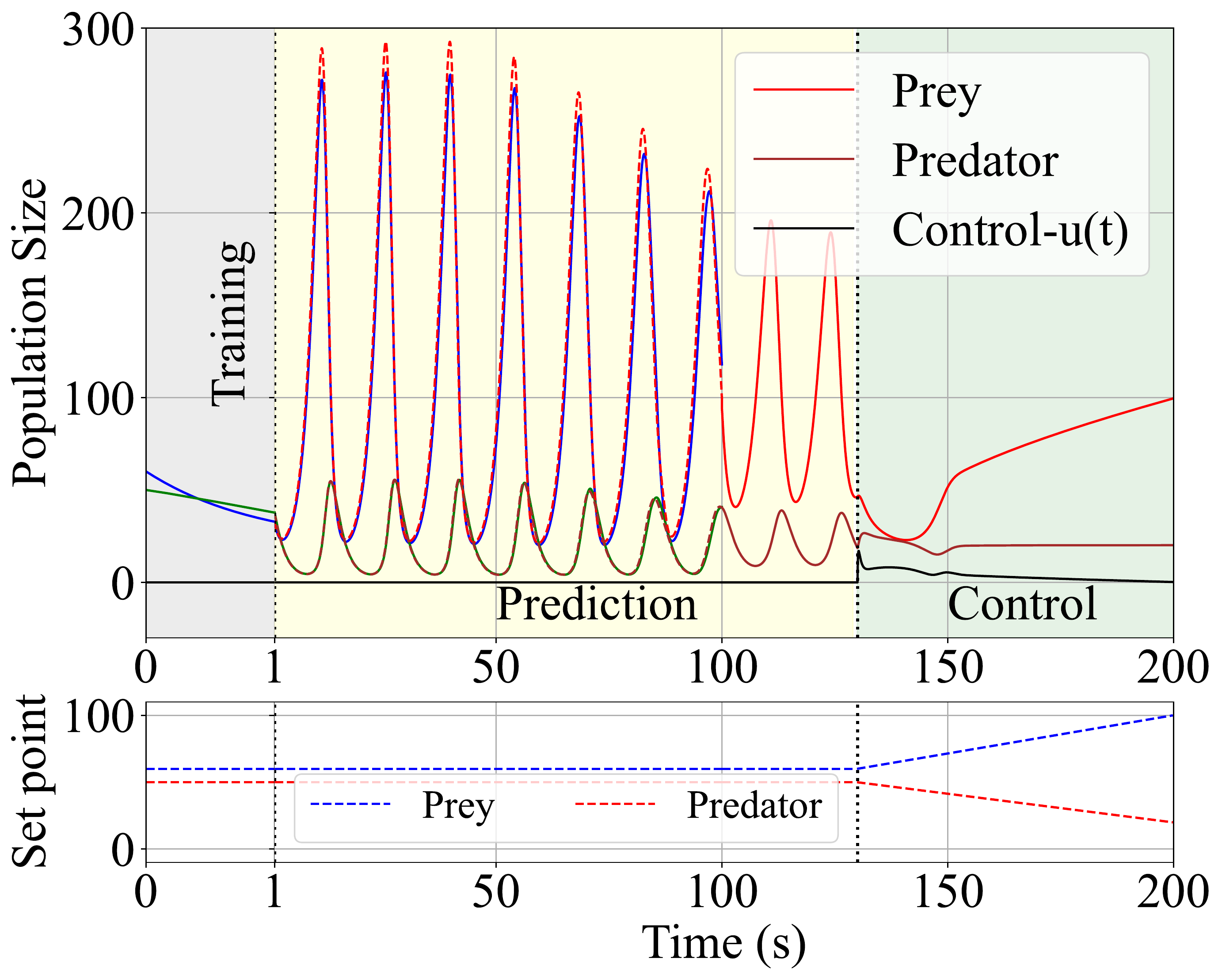} 
        \caption{Performance for evolving set-point}
        \label{fig_lotka_setpoint}  
     \end{subfigure}
     \hfill
     \begin{subfigure}[b]{0.49\textwidth}
         \centering
         \includegraphics[width=\textwidth]{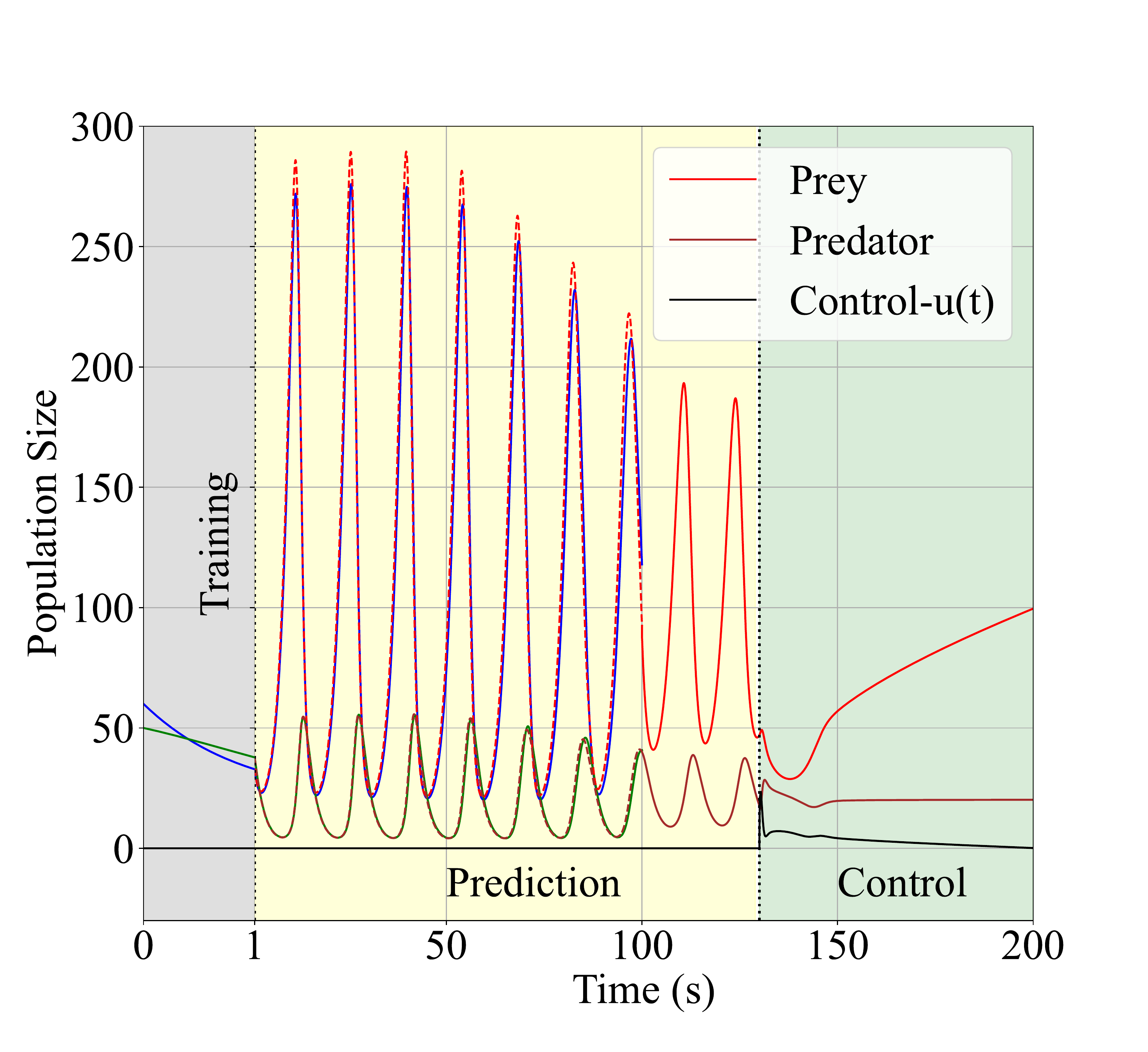} 
        \caption{Performance for time-delayed control input}
        \label{fig_lotka_deadtime} 
     \end{subfigure}
        \caption{Prediction and control performance for changing set-point and time-delayed control input: the absolute relative error is obtained in terms of the noiseless response. In the case study for set-point changes, the reference vector is modeled as a gradually increasing set between \{[60,50], [100,20]\}. In Fig. (b), during the control stage, a time delay of 0.5s is considered. Both (a) and (b) demonstrate the ability of the proposed MASMPC to take into account the changing set-points and dead times.}
        \label{fig:lotka_case_ii}
\end{figure}

\subsubsection{Robustness to adapt with evolving set-points}
In an active control platform, the set points are often subjected to time changes. In order to show that the proposed approach is able to handle the changes in set points and drive the underlying system actively based on the timely evolving set points, we perform a case study by creating a set of $d$ partitions of gradually increasing set points between the reference points $\{ \bm{X}^{crit}_{0}=[60,50]\}$ and $\{[100,20]=\bm{X}^{crit}_{d}\}$. The results are presented in Fig. \ref{fig_lotka_setpoint}. From the results, it is straightforward to comprehend that the proposed framework is well able to handle the continuously evolving set points without any difficulties.

\subsubsection{Ability to control problems with dead-times}
During practical implementations, the performance of active control algorithms is frequently hindered by the time delays between the instant when the control action is executed and the time when the controlled response shows a reaction to the corresponding controller action. Since our proposed MASMPC algorithm essentially works on the principles of the MPC it has the ability to account for future predictions in addition to the present time slot. To demonstrate the performance of the proposed MASMPC algorithm, the same Prey-Predator model is studied by considering a time delay of 0.5s in the control input. The results are encapsulated in Fig. \ref{fig_lotka_deadtime}. As can be seen that the proposed algorithm suffers some convergence issues initially, but with time it essentially drives the system to the reference point. Here, the time taken to reach the reference point can further be shortened by considering more wider prediction horizon.

\begin{figure}[htbp!]
     \centering
     \begin{subfigure}[b]{0.48\textwidth}
         \centering
         \includegraphics[width=\textwidth]{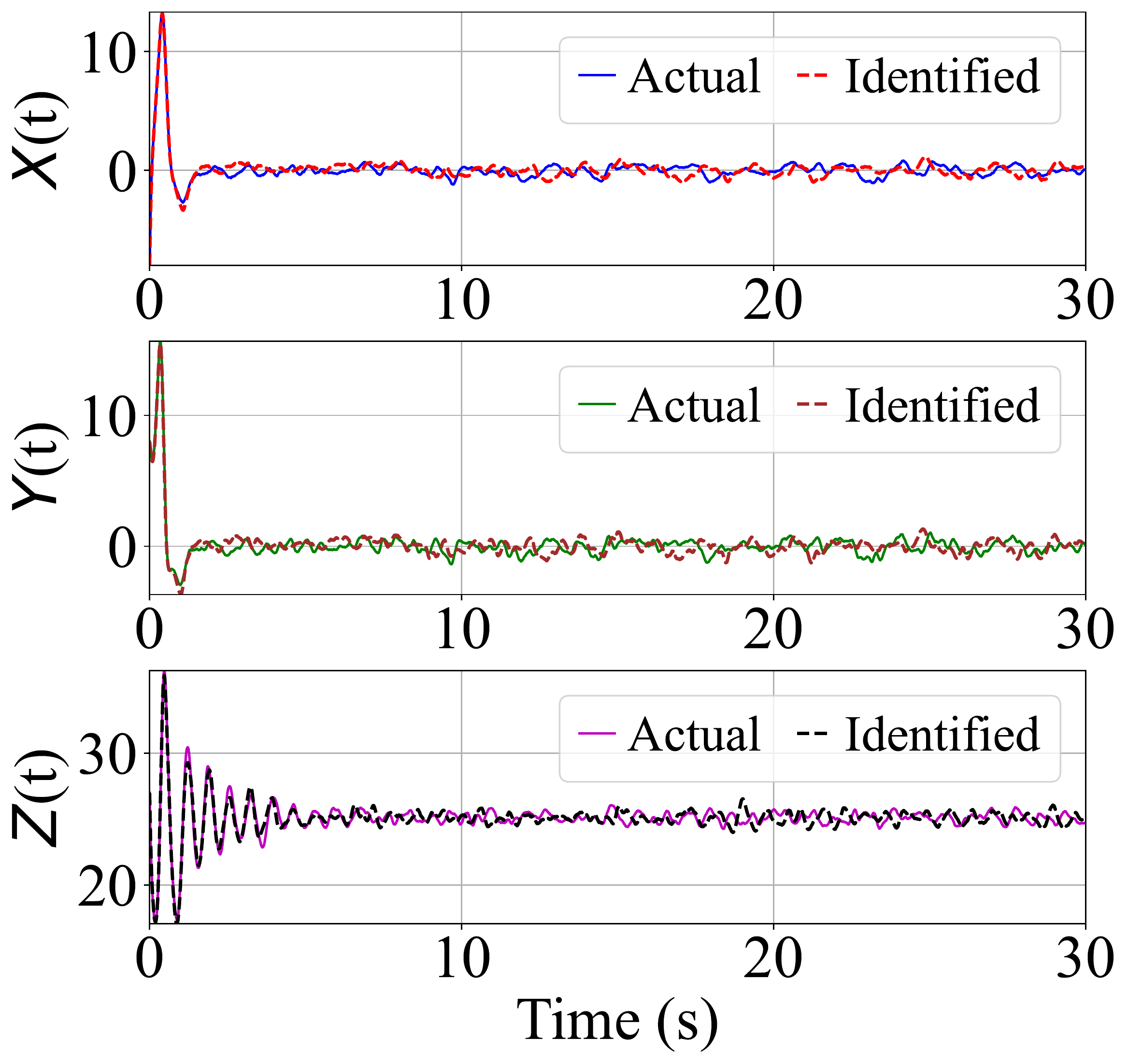}
         \caption{Prediction}
         \label{fig_lorenz1} 
     \end{subfigure}
     \hfill
     \begin{subfigure}[b]{0.51\textwidth}
         \centering
         \includegraphics[width=\textwidth]{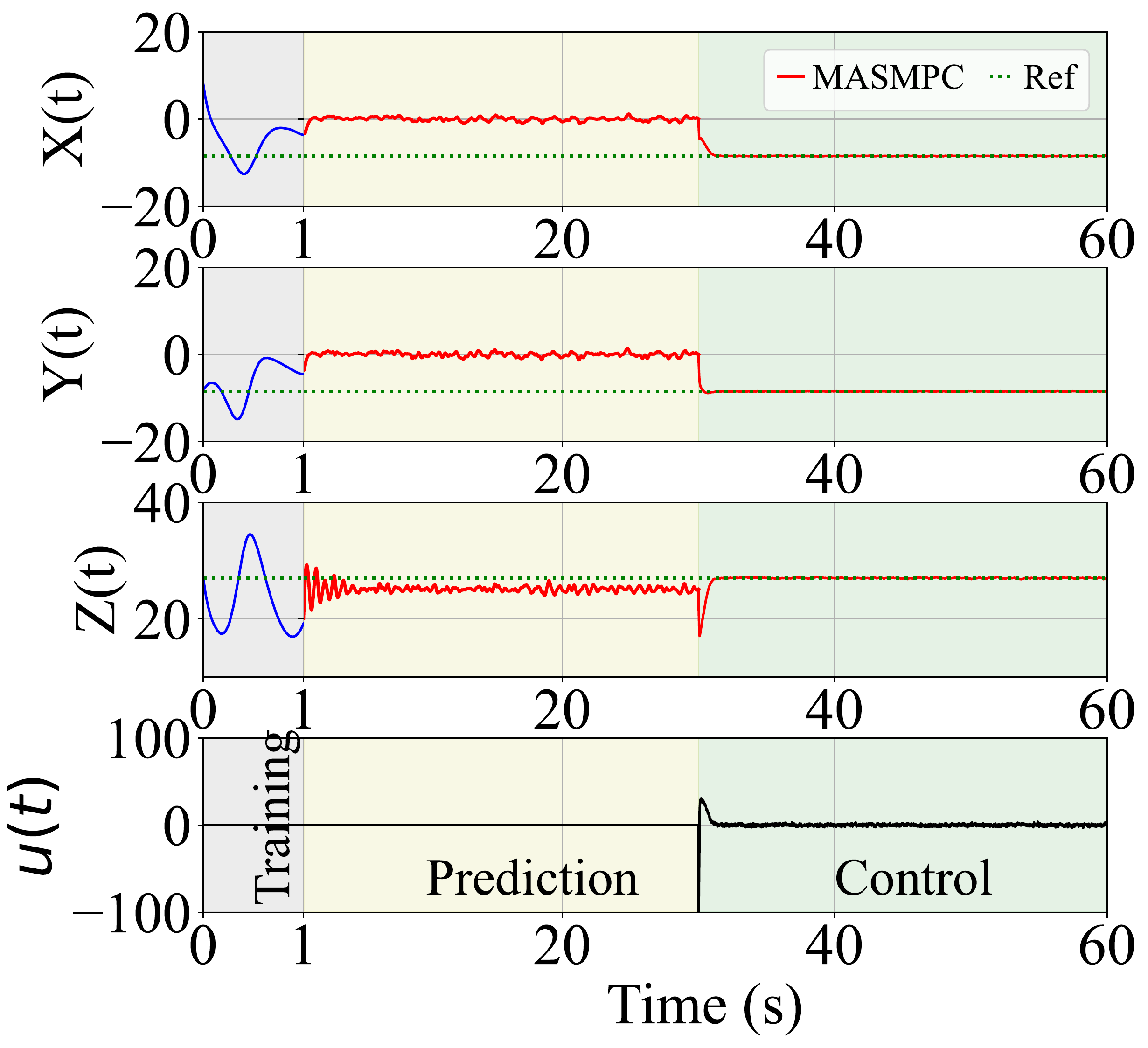} 
         \caption{Control}
         \label{fig_lorenz2}  
     \end{subfigure}
     \caption{Validation and control performance for the Lorenz oscillator: one-second data is used to train the model, after which validation is done for 30 seconds, followed by control from 30$^{th}$ second. Note that the prediction is not explicitly required for control. It is presented here only for validation purposes and visual clarity of the immediate effect of control.}
     \label{fig:lorenz}
\end{figure}

\subsection{Chaotic system modelling: Lorenz oscillator}\label{example2}
The Lorenz system is a nonlinear and non-periodic oscillator that consists of three coupled differential equations. The Lorenz oscillator is highly initial condition sensitive. Given a set of system parameter values, a small perturbation in the initial condition grows very rapidly. Thus the Lorenz oscillator is often used to model chaos in dynamical systems. In its original format, it is a system of deterministic differential equations whose applications can be found in models for dynamos, electric circuits, chemical reactions forward osmosis, etc. Due to the chaotic behavior of the Lorenz oscillator, the identification, prediction, and control of such chaotic systems are challenging \cite{poland1993cooperative,tzenov2014strange}. Further, in the absence of the input measurement, the simultaneous identification and control of chaotic systems may lead to faulty long-term future prediction and control. In order to demonstrate the robustness of the proposed scheme, the Lorenz oscillator is considered as a second example. Since the identification is done without input information, the deterministic Lorenz oscillator is framed as a stochastic system. In a stochastic framework, the controlled Lorenz oscillator can be represented as,
\begin{equation}\label{eq_lorenzsde}
    \begin{aligned}
    &{dX_1} = (\alpha ({X_2} -{X_1})) dt + \sigma_1 dB_1 + u \\
    &{dX_2} = ({X_1}(\rho - {X_3}) - {X_2}) dt + \sigma_2 dB_2 + u \\
    &{dX_3} = ({X_1}{X_2} - \beta {X_3}) dt + \sigma_3 dB_3 + u,
    \end{aligned}
\end{equation}
where $\{ \alpha, \rho, \beta\}$ are the parameters of the system and $u$ is the control input. The solution to the above SDEs is obtained using EM mapping. The covariance of the diffusion matrix is given as,
\begin{equation}
    \Gamma  = \left[ {\begin{array}{*{20}{c}}
    {\sigma _1^2}&0&0\\
    0&{\sigma _2^2}&0\\
    0&0&{\sigma _3^2}
    \end{array}} \right].
\end{equation}
Again the proposed MASMPC algorithm doesn't have any prior knowledge about the underlying physics and only has access to noisy data.
A constraint of the form $-100< u <100$ is applied in the optimization process, which reflects the controller's capacity so that the control force does not exceed the controller's specifications.

For any given arbitrary parameters $\{\alpha, \beta, \rho\}$, the Lorenz oscillator attains a stable solution at the reference points $\{\pm \sqrt{\beta (\rho -1)}, \pm \sqrt{\beta (\rho -1)}, (\rho-1) \} =\{\pm \sqrt{72}, \pm \sqrt{72}, 27 \}$. The control objective is to drive the system to one of these points. The training is similarly performed using one-second data sampled at 1000Hz. The prediction using the trained model is presented in Fig. \ref{fig_lorenz1}. The results are highly promising, as the Lorenz oscillator itself is chaotic, and the presence of external disturbance adds an additional complexity to the identification. However, the proposed framework is able to almost perfectly encapsulate the fundamental physics of such chaotic processes and forecasts a future that is close enough to the original response. In Fig. \ref{fig_lorenz2}, the control performance is plotted. With one-second training, the prediction is performed for 30 seconds, after which the control is turned on. As soon as the control is turned on, the proposed framework drives the solution towards reference, as depicted in Fig. \ref{fig_lorenz2}. This demonstrates the proposed framework's ability to discover, anticipate, and control the correct governing physics of a chaotic and controlled system when disturbance information is unavailable, and data is just one second long. Similar to the previous example, the performance of the proposed framework for this example is compared with the deep learning-based LSTM network in terms of the absolute mean errors, whose results are given in Table. \ref{tab:efficiency}. The results demonstrate that the errors in the results obtained using the proposed scheme are significantly less (almost 10-500 times) than that of LSTM network. Such kind of accuracy in the noisy and low data limit clearly highlights the industrial implementability of the proposed framework.

\subsection{A practical example: control of 76 DOF slender structure}\label{sec:example3}
\begin{figure}
     \centering
     \begin{subfigure}[b]{0.53\textwidth}
         \centering
         \includegraphics[width=\textwidth]{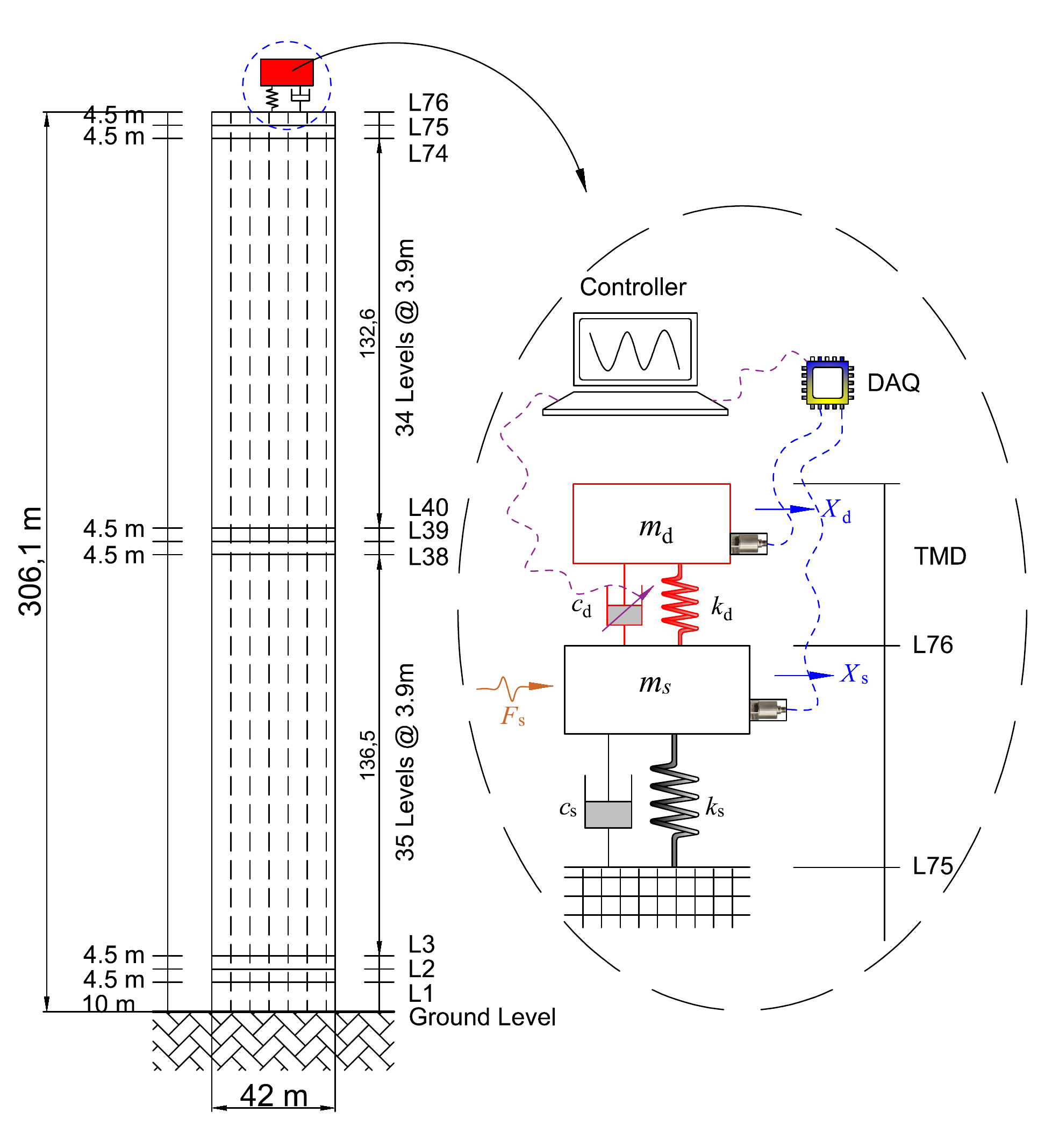} 
        \caption{Elevation of the structure}
        \label{fig_TMD1}
     \end{subfigure}
     \hfill
     \begin{subfigure}[b]{0.46\textwidth}
         \centering
         \includegraphics[width=\textwidth]{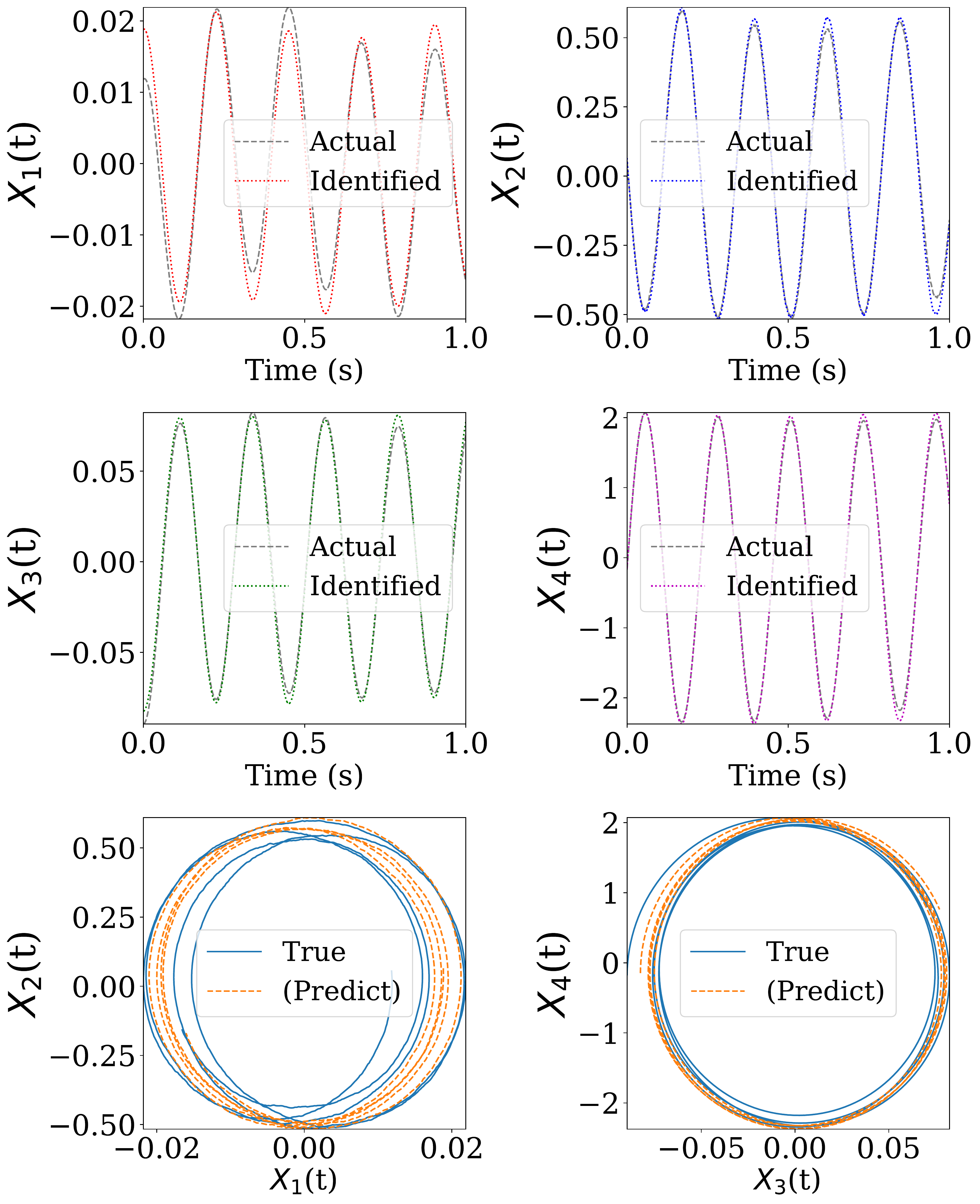} 
        \caption{Prediction performance}
        \label{fig_TMD2}
     \end{subfigure}
     \hfill
     \begin{subfigure}[b]{\textwidth}
         \centering
         \includegraphics[width=\textwidth]{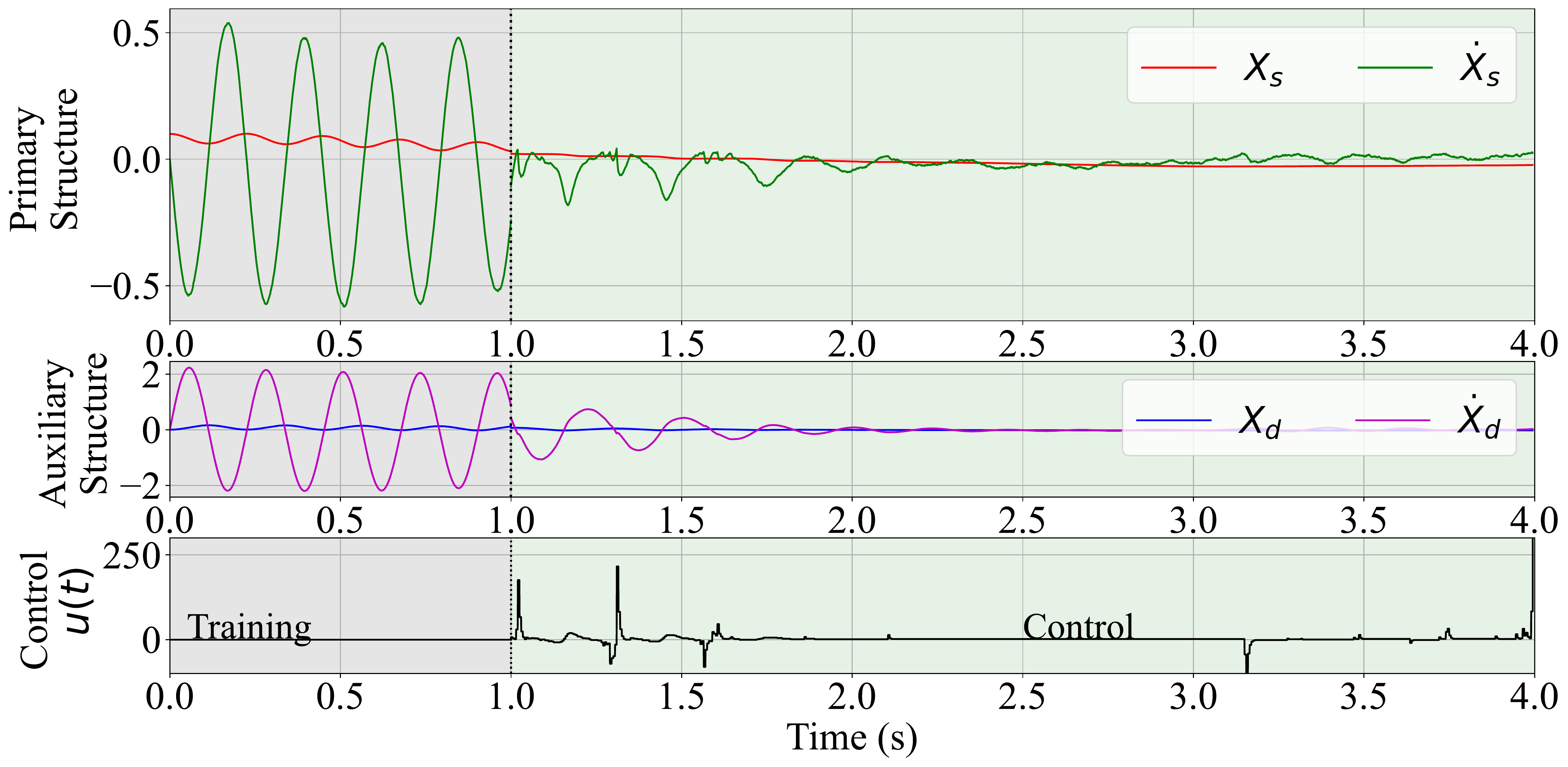} 
        \caption{Control results of the Tuned Mass Damper}
        \label{fig_TMD3}
     \end{subfigure}
        \caption{Validation and control performance for the TMD system: only the vibration of the primary structure is controlled. As a consequence, the TMD responses also get stabilized near the equilibrium solution. The control force here represents the amount of damping to supplied by the shock absorber. Also note that the prediction phase is not a necessity for control in the actual scenario, but is presented here only to visually convey the immediate effect of the control.}
        \label{fig:tmd}
\end{figure}

The applicability of the proposed MASMPC algorithm for real-life problems is demonstrated in this section using a realistic structure. The undertaken structure was originally proposed for the city of Melbourne, Australia, and later proposed as a benchmark for response control in Ref. \cite{yang2004benchmark}. The 76-story structure is a reinforced concrete office tower having a height of 306.1m and a base width of 42m, creating a slenderness ratio of 7.3. This makes the structure sensitive to wind excitation. To control the response of the structure, a semi-active tuned mass damper (SATMD) using the linear quadratic Gaussian (LQG) approach is developed in the literature \cite{yang2004benchmark}. In this work, we propose to model the SATMD with a controller that is devised using the proposed MASMPC algorithm. In the proposed SATMD, we try to tune the tuned mass damper by continuously tuning the optimal damping constant of the mass damper in real-time, thereby achieving active control of the vibrations of the primary structure. In practice, this strategy can be implemented by using magnetorheological shock absorbers, which are filled with magnetorheological fluid. This special type of fluids can be controlled by a magnetic field \cite{poynor2001innovative}, and therefore, by varying the power of the electromagnet, the damping characteristics of the shock absorber can be continuously controlled. A similar framework was proposed for the semi-control of structures by controlling the damping parameter of the TMD using an energy-based predictive (EBP) algorithm \cite{zelleke2019semi}. The SATMD cum controller is devised on the top floor level of the 76-story benchmark. The elevation and the proposed SATMD module are depicted in Fig. \ref{fig_TMD1}. For more information on the benchmark structure, the readers are referred to Ref. \cite{yang2004benchmark}.

While the wind force data on the benchmark structure were determined from wind tunnel experiments on a scaled model, in this work the wind force is assumed as non-measurable and modeled as Brownian motions applied at all floor levels. The governing physics of the SATMD along with the benchmark structure subjected to random Brownian motions in the floor levels is given by the following equation,
\begin{equation}
    {{\bf M}_s}{{\ddot {\bm X}}_s} + {{\bf C}_s}{{\dot {\bm X}}_s} + {{\bf K}_s}{{\bm X}_s} = {\bf G}{\dot {\bm B}},
\end{equation}
where ${\bf M}$, ${\bf K}$ and ${\bf C}$ are the $\mathbb{R}^{77 \times 77}$ mass, stiffness, and damping matrices, which can be obtained from Ref. \cite{yang2004benchmark}; ${\bm X}=[X_1, \ldots, X_{76}, X_d]$ is the response vector with $X_d$ being the displacement of the mass damper; ${\bf G} \in \mathbb{R}^{77 \times 77}$ is the force intensity matrix, in this case, a diagonal matrix; and ${\bm B} \in \mathbb{R}^{77}$ is the wind excitation vector modeled as Brownian motions. To define the problem more clearly, let us write the equation of motion of 76$^{th}$ and TMD floor,
\begin{equation}\label{eq_tmd}
    \begin{aligned}
        \left[ {\begin{array}{*{20}{c}}
        {{M_{76}}}&0\\
        0&{{M_d}}
        \end{array}} \right]\left[ {\begin{array}{*{20}{c}}
        {{{\ddot X}_{76}}}\\
        {{{\ddot X}_d}}
        \end{array}} \right] + \left[ {\begin{array}{*{20}{c}}
        { - {C_{76}}}&{{C_{76}} + {{\tilde{C}}_d}}&{ - {{\tilde{C}}_d}}\\
        0&{-{{\tilde{C}}_d}}&{{{\tilde{C}}_d}}
        \end{array}} \right]\left[ {\begin{array}{*{20}{c}}
        {{{\dot X}_{75}}}\\
        {{{\dot X}_{76}}}\\
        {{{\dot X}_d}}
        \end{array}} \right] + \\
        \left[ {\begin{array}{*{20}{c}}
        { - K_{76}}&{{K_{76}} + {K_d}}&{ - {K_d}}\\
        0&{-{K_d}}&{{K_d}}
        \end{array}} \right]\left[ {\begin{array}{*{20}{c}}
        {{X_{75}}}\\
        {{X_{76}}}\\
        {{X_d}}
        \end{array}} \right] = \left[ {\begin{array}{*{20}{c}}
        {{\sigma _{76}}}&0\\
        0&0
        \end{array}} \right]\left[ {\begin{array}{*{20}{c}}
        {{{\dot B}_{76}}}\\
        {0}
        \end{array}} \right].
    \end{aligned}
\end{equation}
Here, ${\tilde{C}_d} = {C_d}+u(t)$ is the tuned damping constant of the mass damper, and the objective is to obtain the control parameter $u(t)$.
In view of computational, instrumentation, and maintenance cost effectiveness we further assume that the sensors are placed only on the 76 DOF and TMD levels. Thus we model the problem as a partially observed system and try to control the vibration of the slender benchmark structure arising due to the wind force using only the output measurements at those two floors. With an appropriate statespace transformation, the corresponding It\^{o}-stochastic SDEs for the dynamical system are derived as Eq. \eqref{sdeg}, and then solved using the EM scheme.
The simulation parameters for the controller are provided in Table. \ref{table_param}. Two different timescales are used in prediction and control, which are ${\Delta} t$=0.0005 for prediction and $\tilde{\Delta} t$=0.01 for the control algorithm. The aim of the control in this problem is to stabilize the primary system such that the corresponding solutions remain near the vicinity of the steady state solution $\{X_s, \dot{X}_s\}^{\rm{steady}}=\{0, 0\}$. Following additional constraints are applied to the optimization, (i) the velocity component at the top floor does not exceed 0.05 m/s and (ii) at any time the external damping to be supplied does not exceed 2000 N-s/m, where the latter is indicative of the capacity of the absorber.
\begin{equation}
    \begin{array}{c}
        -0.05 < {\dot{X}}_{76} < 0.05 \\
        -2000 < u < 2000.
    \end{array}
\end{equation}

Similar to previous problems, the training is performed using one-second data only and for learning the stochastic model only the output responses of 76 DOF and the mass damper are utilized. Afterward, the validation is carried out as depicted in Fig. \ref{fig_TMD2}. The results show complete overlapping of the solutions obtained using the identified actual model. The control results obtained using the discovered model are provided in Fig. \ref{fig_TMD3}. Once the control is turned on, the proposed framework drastically drives the primary structure response to the reference solution. Because the goal is not to control the mass damper vibration, the responses of the auxiliary structure do not instantly reflect the effect of control; however, as the primary system's response decreases, the mass damper follows the same path. Overall, through these examples, we demonstrated the salient features of the proposed MASMPC. This includes (i) no information on underlying physics, (ii) non-dependency on the external input information, (iii) good performance at a low data limit, only one-second, (iv) probabilistic learning, (v) non-conservative control, (vi) ability to adapt with set-point changes, (vii) ability to account multiple variables, (viii) ability to handle dead-time, and (ix) robustness to discover the physics of partially observed systems.

\section{Conclusions}\label{sec:conclusion}
We propose a novel model agnostic stochastic model predictive control (MASMPC) framework in this paper. The role of model identification is immense in any model-free control algorithm. Traditionally, this is achieved by using black-box type machine learning or deep learning models; however, such models usually do not generalize to the unseen environment and hence, one has to continuously update the identified model by repeatedly invoking the training loop. This is one of the major bottlenecks of model predictive control and jeopardizes the real-time applicability of model predictive control algorithms. We herein propose a novel approach that leverages Bayesian statistics and Ito calculus to learn the governing differential equation of the system. With such a setup, the learned model is interpretable and seamlessly generalizes to an unseen environment. This eliminates the need for repeated retraining, and the same model (learned) can be used perpetually. Additionally, from a practical consideration, we often only have access to noisy measurements of the state variables with no information/measurements on the input force. The proposed MASMPC framework is tailormade to address this issue as it learns the governing physics in the form of a stochastic differential equation with the forcing part modeled as white Gaussian noise. Finally, the proposed approach (a) can seamlessly tackle set-point change, (b) works with multiple control variables, and (c) is robust against measurement noise. 
The proposed framework is applied to three nonlinear oscillators that are utilized for modeling periodic and chaotic behaviors of various processes in the stock market, biology, ecosystem, chemical, and mechanical sciences. The results are promising and show accurate identification of the actual process dynamics without requiring the input force information and successful estimation of the correct control force.

In terms of the applications, the proposed MASMPC approach uses only one second of data sampled at a frequency of 1000Hz. Obtaining this amount of data is no longer a limitation thanks to recent advances in sensor development. Because the model identified using the proposed scheme directly describes the governing physics of the underlying process, it can account for the dramatic shift in system dynamics. The identification of the controlling physics from relatively fewer amounts of data, without the need for input force information, permits flawless integration with the MPC framework. Further, a common issue that frequently arises in the deployment of control algorithms for industrial use is dead time. In control, the dead time refers to the time required by the controlled system to react in response to the control force supplied by the mechanical controller device. However, it can be recalled that the proposed framework tries to find an optimal control force based on the predictions made over a finite prediction horizon. Thus, the proposed SMPC framework provides a remedy to avoid the issue of dead time by appropriately choosing the prediction horizon length based on the data obtained from the time when the proposed framework is deployed for the first time. To summarize, when the input information is unavailable and the output measurement data is limited, this framework offers a robust algorithm for real-time control of non-linear and chaotic systems.

In this work, only a handful of examples and associated case studies are presented in order to demonstrate the potential applicability of the proposed MASMPC for industrial applications. However, for industrial and broad-level applications, certain improvements are required to be incorporated into the proposed MASMPC framework as a future extension of this work. For example, natural systems are often described using partial differential equations (PDEs). In such cases, the extension of the proposed framework (i) to discover PDEs without the input measurement in terms of stochastic PDEs (SPDEs) and (ii) to devise a control strategy for controlling the systems described using SPDEs needs to be carried out independently, which are beyond the scope of the present work. The proposed MASMPC framework should further be investigated for the non-stationary types of excitation, which is a common phenomenon for earthquake and wind exciting systems. As an initial thought, this can be done by employing the relations between Brownian motion and Kanai-Tajimi filters, however, this needs to be investigated in a separate work.

\appendix

\appendix

\section{Kramers-Moyal expansion for estimation of the drift and diffusion terms of an SDE from sample paths}\label{appenA}
Let us consider, $p(X, t)$=${\rm{P}}(X,t|X_0,t_0)$ to be the transition probability density of the solution of the SDE in Eq. \eqref{sdeg}. Then the Kramers-Moyal expansion is written as \cite{risken1996fokker}:
\begin{equation}\label{A1}
    \frac{{\partial P(X,t)}}{{\partial t}} = {\sum\limits_{n = 1}^\infty  {\left( { - \frac{\partial }{{\partial X}}} \right)} ^n}{D^{(n)}}(X,t)p(X,t),
\end{equation}
where the coefficients in the expansion are given as:
\begin{equation}
    {D^{(n)}}(X) = {\left. {\frac{1}{{n!}}\mathop {\lim }\limits_{{\Delta t}  \to 0} \frac{1}{{\Delta t} }\left\langle {{{\left| {X(t + {\Delta t} ) - z} \right|}^n}} \right\rangle } \right|_{X(t) = z}}.
\end{equation}
In order to know how many terms in Eq. \eqref{A1} will be active for the SDE in Eq. \eqref{sdeg}, one can refer the Fokker-Planck equation for the pdf of the solution of Eq. \eqref{sdeg}, given as,
\begin{equation}
    \begin{aligned}
    \frac{{\partial p(X,t)}}{{\partial t}} =  - \frac{{\partial (p(X,t)f(X,t))}}{{\partial X}} + \frac{1}{2}\frac{{{\partial ^2}(p(X,t){g^2}(X,t))}}{{\partial {X^2}}}; \quad p(0,X) = {p_0}(X),
    \end{aligned}
\end{equation}
where the random variable $X$ satisfies the SDE in Eq. \eqref{sdeg}. At this point it is clear that there will be two terms active in  Eq. \eqref{sdeg}. Thus for $n$=2, Eq. \eqref{A1} yields,
\begin{equation}\label{A10}
    \frac{{\partial P(X,t)}}{{\partial t}} =  - \frac{\partial }{{\partial X}}\left[ {{D^{(1)}}(X,t)p(X,t)} \right] + \frac{{{\partial ^2}}}{{\partial {X^2}}}\left[ {{D^{(2)}}(X,t)p(X,t)} \right].
\end{equation}
On comparison of Eqs. \eqref{A1} and \eqref{A10}, it is straightforward to note that to estimate the drift and diffusion terms it suffices to estimate the first and second order moments of the variations of the random variable $X(t)$ and then calculate the coefficients ${D^{(1)}}$ and ${D^{(2)}}$, formally, $f(X,t) = {D^{(1)}}$ and $g^2(X,t) = {D^{(2)}}$. To derive these coefficients in the Kramers-Moyal expansion, it is imperative to understand the one-step It\^{o}-Taylor expansion of the random variable $X(t)$. Let us consider the integral form the It\^{o}-lemma \cite{tripura2020ito}:
\begin{equation}
    F({X_{t + h}},t + h) = F({X_t},t) + \int_t^{t + h} {{\Im ^0}\left( {F({X_s},s)} \right)ds } + \int_t^{t + h} {{\Im ^1}\left( {F({X_s},s)} \right)dB(s)},
\end{equation}
where the operators ${\Im ^0}(.)$ and ${\Im ^1}(.)$ are given as,
\begin{equation}\label{operator}
    \begin{array}{ll}
    {\Im ^0}(.) &= \frac{{\partial (.)}}{{\partial t}} + \sum\limits_i^m {{f_i}({X_t},t)\frac{{\partial (.)}}{{\partial {X_i}}}}  + \frac{1}{2}\sum\limits_i^m {\sum\limits_j^m {\sum\limits_k^n {{g_{i,k}}({X_t},t){g_{j,k}}({X_t},t)\frac{{{\partial ^2}(.)}}{{\partial {X_i}\partial {X_j}}}} } } \\
    {\Im ^1}(.) &= \sum\limits_i^m {\sum\limits_k^n {{g_{i,k}}({X_t},t)\frac{{\partial (.)}}{{\partial {X_i}}}} }.
    \end{array}
\end{equation}
Then substituting $F({X_t},t) = X(t)$, one verifies that ${\Im ^0}X(t) = f({X_t},t)$, and ${\Im ^1}X(t) = g({X_t},t)$. This yields the first iteration:
\begin{equation}\label{iteration1}
    \begin{array}{ll}
     X(t + h) &= X(t) + \int_t^{t + h} {{\Im ^0}\left( {X(s)} \right)ds + } \int_t^{t + h} {{\Im ^1}\left( {X(s)} \right)dB(s)} \\
     &= X(t) + \int_t^{t + h} {f({X_s},s)ds + } \int_t^{t + h} {g({X_s},s)dB(s)}.
    \end{array} 
\end{equation}
In order to perform the second iteration it is required to find the stochastic expansion of the terms $F({X_t},t) = f({X_s},s)$, and $F({X_t},t) = g({X_s},s)$. Thus, expanding $f({X_t},t)$, and $g({X_t},t)$ and using the operators in Eq. \eqref{operator} yields,
\begin{equation}
\begin{array}{ll}
f({X_s},s) =& f(X,t) + \int_t^{{s_1}} {{\Im ^0}\left( {f({X_{{s_2}}},{s_2})} \right)d{s_2} } + \int_t^{{s_1}} {{\Im ^1}\left( {f({X_{{s_2}}},{s_2})} \right)dB({s_2})} \\
g({X_s},s) =& g(X,t) + \int_t^{{s_1}} {{\Im ^0}\left( {g({X_{{s_2}}},{s_2})} \right)d{s_2} } + \int_t^{{s_1}} {{\Im ^1}\left( {g({X_{{s_2}}},{s_2})} \right)dB({s_2})}.
\end{array}
\end{equation}
On substituting the above result in Eq. \eqref{iteration1} and further iterating,
\begin{equation}\label{final_itotaylor}
\begin{array}{ll}
    X(t + h) - X(t) = & f(X,t)\int_t^{t + h} {d{s_1} + } g(X,t)\int_t^{t + h} {dB({s_1})} + \\
    & {\Im ^1}\left( {g(X,t)} \right)\int_t^{t + h} {\int_t^{{s_1}} {dB({s_2})dB({s_1})} }  + R,
\end{array}
\end{equation}
where the remainder term $R$ contains the higher order multiple stochastic integrals. It can be noticed that the above equation is infinitely expandable using the Eq. \eqref{operator}. For further treatment, the first moment is taken on both side. After taking the first moment on both sides and noting the results, (i) $\left\langle {\int_t^{t + h} {dB({s_1})} } \right\rangle  = 0$ and (ii) $\left\langle {\int_{{t_0}}^t {\int_{{t_0}}^{{s_1}} {dB({s_2})dB({s_1})} } } \right\rangle  = \left\langle {\int_{{t_0}}^t {B({s_1})dB({s_1})}  - B(t)\int_{{t_0}}^t {dB({s_1})} } \right\rangle = 0$, the following is obtained,
\begin{equation}
    \begin{array}{l}
    \left\langle {X(t + h) - X(t)} \right\rangle  \\
    = f(X,t)h + g(X,t)\left\langle {\int_t^{t + h} {dB({s_1})} } \right\rangle  + {\Im ^1}\left( {g(X,t)} \right)\left\langle {\int_t^{t + h} {\int_t^{{s_1}} {dB({s_2})dB({s_1})} } } \right\rangle  + \left\langle R \right\rangle \\
    = f(X,t)h + \left\langle R \right\rangle .
    \end{array}
\end{equation}
If the number of iteration is denoted by $k$, then the multiple Brownian integrals (with multiplicity $k$) of the type $\int_t^{t + h} {\int_t^{{s_1}} { \ldots \int_t^{{s_k}} {dB({s_{k + 1}}) \ldots } dB({s_2})dB({s_1})} }$ of multiplicity $k$ have a contribution proportional to $h^{k/2}$, the time integrals $\int_t^{t + h} {\int_t^{{s_1}} { \ldots \int_t^{{s_k}} {d{s_{k + 1}} \ldots } d{s_2}d{s_1}} } $ have a contribution proportional to $h^k$, and the combination  shares a contribution between $h^{k/2}$ and $h^{k}$. Thus it is easy to infer that as $h \to 0$, the higher order terms in the remainder $R$ will vanish. After invoking this facts in the above expression, the first coefficient in Kramers-Moyal expansion is obtained as,
\begin{equation}
    {D^{(1)}} = f(X,t) = {\left. {\mathop {\lim }\limits_{\Delta t \to 0} \frac{1}{{\Delta t}}\left\langle {{X^{(1)}}(t + \Delta t) - z} \right\rangle } \right|_{X(t) = z}}.
\end{equation}
To derive the second coefficient, it is only required to find the quadratic variation of the increment process of $X(t)$. Upon taking second moment on both sides of Eq. \eqref{final_itotaylor}, the following is obtained,
\begin{equation}
    \begin{array}{ll}
    \left\langle {{{\left| {X(t + h) - X(t)} \right|}^2}} \right\rangle  &= f(X,t)h + g(X,t)\Delta B + \left\langle R \right\rangle \\
     &= {f^2}(X,t){h^2} + 2f(X,t)g(X,t)h\Delta B + {g^2}(X,t){\left( {\Delta B} \right)^2} + \left\langle R \right\rangle \\
     &= {g^2}(X,t)h + \left\langle R \right\rangle .
    \end{array}
\end{equation}
In the above proof, the It\^{o} identities  are utilized, which states that under the mean square convergence theory the quadratic variation of the time and Brownian increments are given as $d{s_2}d{s_1} = 0$, $d{s_1}dW({s_1}) = 0$, $dW({s_2})d{s_1} = 0$, $dW({s_1})dW({s_2}) = 0$. The deterministic time integrals will vanish automatically since they have finite variance and zero quadratic covariance and the other higher order Brownian integrals will vanish as $h \to 0$. Thus the second coefficient in the Kramers-Moyal expansion is obtained as,
\begin{equation}
    {D^{(2)}} = g^2(X,t) = {\left. {\frac{1}{2}\mathop {\lim }\limits_{\Delta t \to 0} \frac{1}{{\Delta t}}\left\langle {{{\left| {{X^{(1)}}(t + \Delta t) - z} \right|}^2}} \right\rangle } \right|_{X(t) = z}}.
\end{equation}
From the ergodic assumption of the evolution of process $X(t)$, the time average $\left\langle \cdot \right\rangle$ is often replaced by the expectation operator $\rm{E}(\cdot)$. For an $m$-dimensional diffusion process the Kramers-Moyal exapansion can be generalised as,
\begin{equation}
    \begin{array}{ll}
    {f_i}({\bm{X}},t) &= {\left. {\mathop {\lim }\limits_{\Delta t \to 0} \frac{1}{{\Delta t}}E\left[ {{X_i}(t + \Delta t) - {z_i}} \right]} \right|_{{X_k}(t) = {z_k}}}\\
    {{\bf{\Gamma}} _{ij}}({\bm{X}},t) &= \Biggl. \frac{1}{2}\mathop {\lim }\limits_{\Delta t \to 0} \frac{1}{{\Delta t}}E\Bigl[ \left| {{X_i}(t + \Delta t) - {z_i}} \right| \left| {{X_j}(t + \Delta t) - {z_j}} \right| \Bigr] \Biggr|_{{X_k}(t) = {z_k}} \quad \forall \;k = 1,2, \ldots m,
    \end{array}
\end{equation}
where ${{\bf{\Gamma}} _{ij}}({\bm{X}},t)$ is the ${ij}^{th}$ element of the covariation matrix ${\bf{\Gamma}} ({{\bm{X}}_t},t) = {\bm{g}}({{\bm{X}}_t},t) {\bm{g}}{({{\bm{X}}_t},t)^T}$.

\section*{Acknowledgements} 
T. Tripura acknowledges the financial support received from the Ministry of Education (MoE), India, in the form of the Prime Minister’s Research Fellowship (PMRF). S. Chakraborty acknowledges the financial support received from Science and Engineering Research Board (SERB) via grant no. SRG/2021/000467 and seed grant received from IIT Delhi.

\section*{Declarations}

\subsection*{Funding} The corresponding author received funding from IIT Delhi in form of seed grant.

\subsection*{Conflicts of interest} The authors declare that they have no conflict of interest.

\subsection*{Availability of data and material} The datasets generated during and/or analysed during the current study are available from the corresponding author on reasonable request.

\subsection*{Code availability} The Python codes written for this work are available from the corresponding author on reasonable request.


\end{document}